\documentclass[aps,prl,twocolumn,superscriptaddress,10pt]{revtex4-2}

\usepackage{graphicx}
\usepackage{dcolumn}
\usepackage{bm}
\usepackage{hyperref}
\usepackage{siunitx}
\usepackage{amsmath, amsfonts}
\usepackage[normalem]{ulem}
\usepackage{xcolor}

\newcommand{\comments}[1]{}   

\DeclareMathOperator{\sgn}{sgn}

\begin{document}


\title{Universal Coarsening and Giant-Cluster Formation in Growing Interfaces}



\author{Renan A. L. Almeida}
\email[]{ra.lisboaalmeida@gmail.com}
\homepage[]{https://sites.google.com/view/renan-almeida/home}
\affiliation{Sorbonne Université, Laboratoire de Physique Théorique et Hautes Energies, CNRS UMR 7589, 4 Place Jussieu, 75252 Paris Cedex 05, France}
\affiliation{Instituto de F\'\i sica, Universidade Federal do Rio Grande do Sul, CP 15051, 91501-970, Porto Alegre RS, Brazil}

\author{Tiago J. Oliveira}
\affiliation{Departamento de F\'\i sica, Universidade Federal de Viçosa, 36570-900, Viçosa MG, Brazil}

\author{Jeferson J. Arenzon}
\affiliation{Instituto de F\'\i sica, Universidade Federal do Rio Grande do Sul, CP 15051, 91501-970, Porto Alegre RS, Brazil}

\author{Leticia F. Cugliandolo}
\affiliation{Sorbonne Université, Laboratoire de Physique Théorique et Hautes Energies, CNRS UMR 7589, 4 Place Jussieu, 75252 Paris Cedex 05, France}


\date{\today}

\begin{abstract}
Clusters formed by fluctuations of two-dimensional (2D) directed interfaces around a threshold level have been extensively studied at equilibrium and in nonequilibrium steady states, but their coarsening dynamics remain poorly understood. 
Here, we numerically investigate this unexplored coarsening of clusters in 2D growing interfaces believed to belong to the Kardar–Parisi–Zhang universality class. 
Using a two-point spatial correlator, we demonstrate statistical time invariance of the evolving configurations and identify scaling forms shared across distinct models. 
We reveal a pronounced asymmetry in the growth of the largest clusters: one cluster emerges as a giant structure whose characteristic length exceeds the correlation length. 
Population-dependent scaling forms for the number densities of cluster areas are uncovered. 
These findings highlight new universal aspects of growing interfaces and suggest avenues for experimental verification.

\end{abstract}


\maketitle




%
Universality, a fundamental concept in modern science, 
refers to the fact that microscopically different interacting systems can exhibit the same large-scale behavior~\cite{Goldenfeld_book18}.
%
Originally discovered for the exceptional condition of critical equilibrium and
elegantly rationalized via the renormalization group~\cite{Goldenfeld_book18},
it has been later recognized to hold for a broad scope of nonequilibrium macroscopic systems 
endowed with dynamic scale invariance~\cite{Bray94, HHealy95, Henkel_book08, *Henkel_book11}.

Many such nonequilibrium systems bear a description in terms of clusters displaying domain growth, also 
called coarsening.
Besides simple magnetic and opinion-formation models \cite{Arenzon07, Dornic01}, 
coarsening occurs in several real systems, 
including
liquid crystals~\cite{Sicilia08, Almeida23, *Almeida25},
superconductors~\cite{Prozorov08},
dry liquid foams~\cite{Lambert10},
and dense bacterial populations~\cite{Mcnally17}.
At late times, 
key features of the intricate set of clusters become statistically time-invariant
when observed through dimensionless quantities involving a single 
time-dependent characteristic length $\xi(t)$.
This property is manifested in the scaling forms of
correlation functions~\cite{Bray94}, 
geometric aspects of clusters~\cite{Sicilia07, *Sicilia09}, and
their number densities~\cite{Arenzon07, Sicilia07, *Sicilia09}, among other observables.
Such forms, supplemented by scaling exponents, are hallmark signatures of the corresponding dynamic universality class.

    
    \begin{figure}[!t]
    \includegraphics[width=0.157\textwidth]{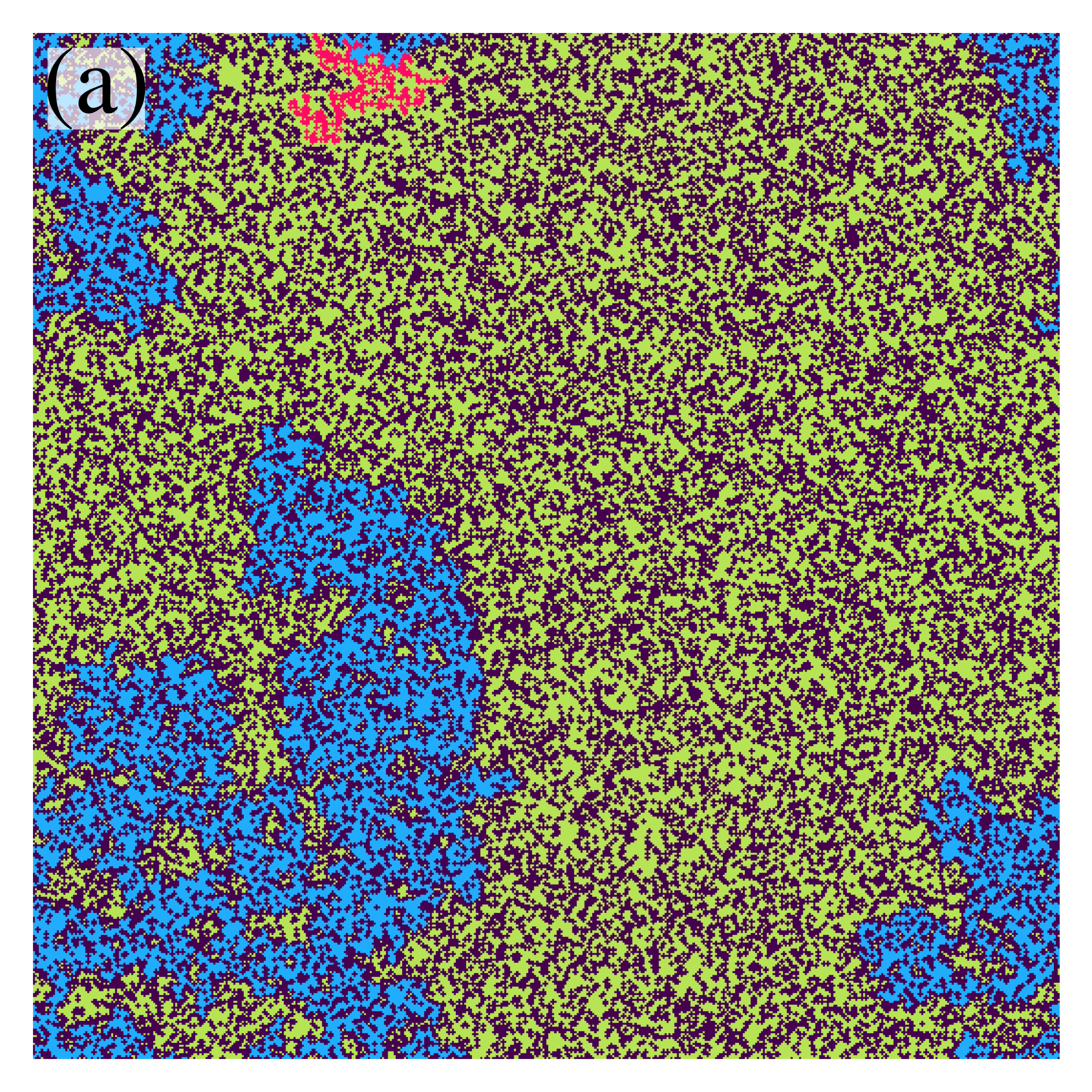}
    \includegraphics[width=0.157\textwidth]{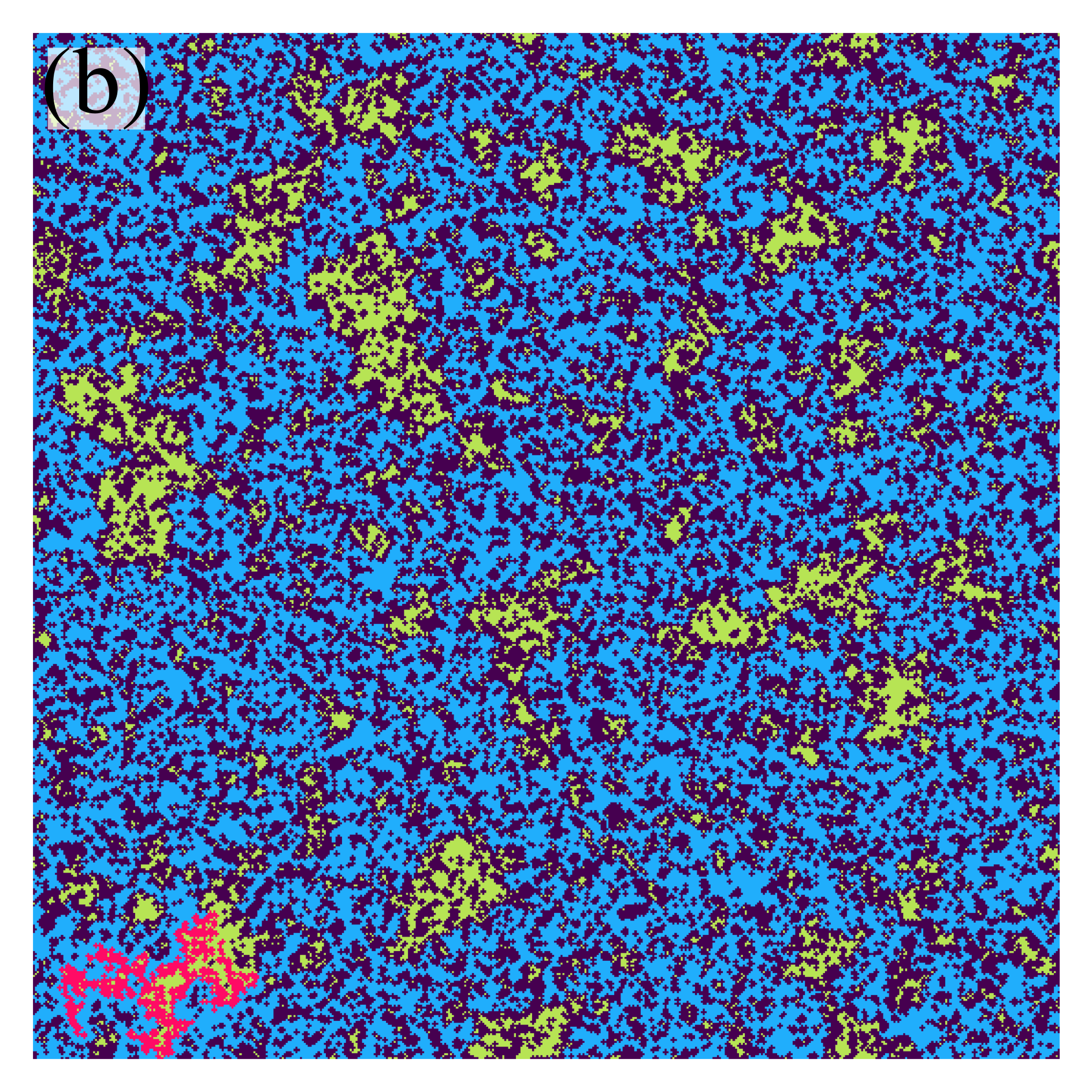}
    \includegraphics[width=0.157\textwidth]{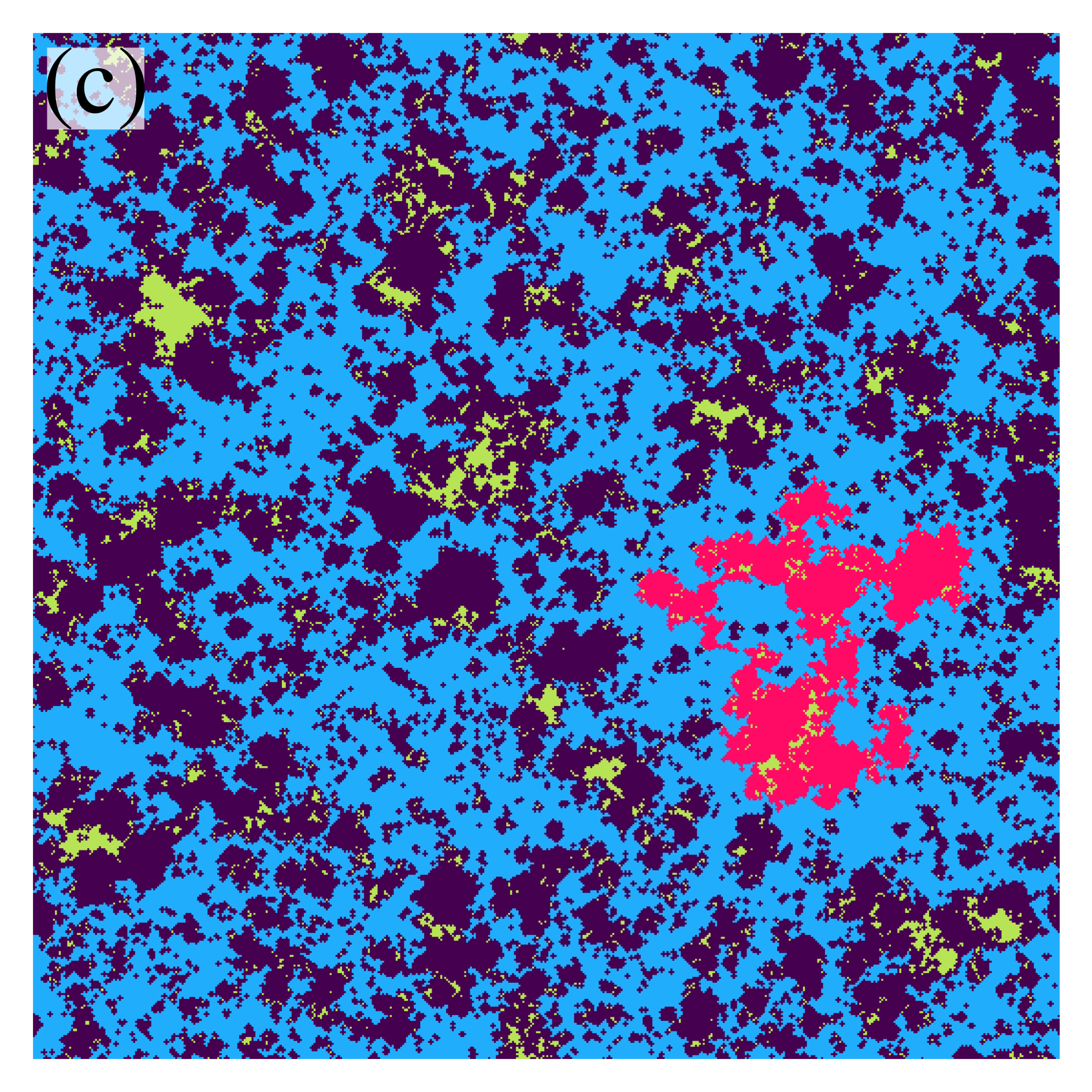}
    \includegraphics[width=0.157\textwidth]{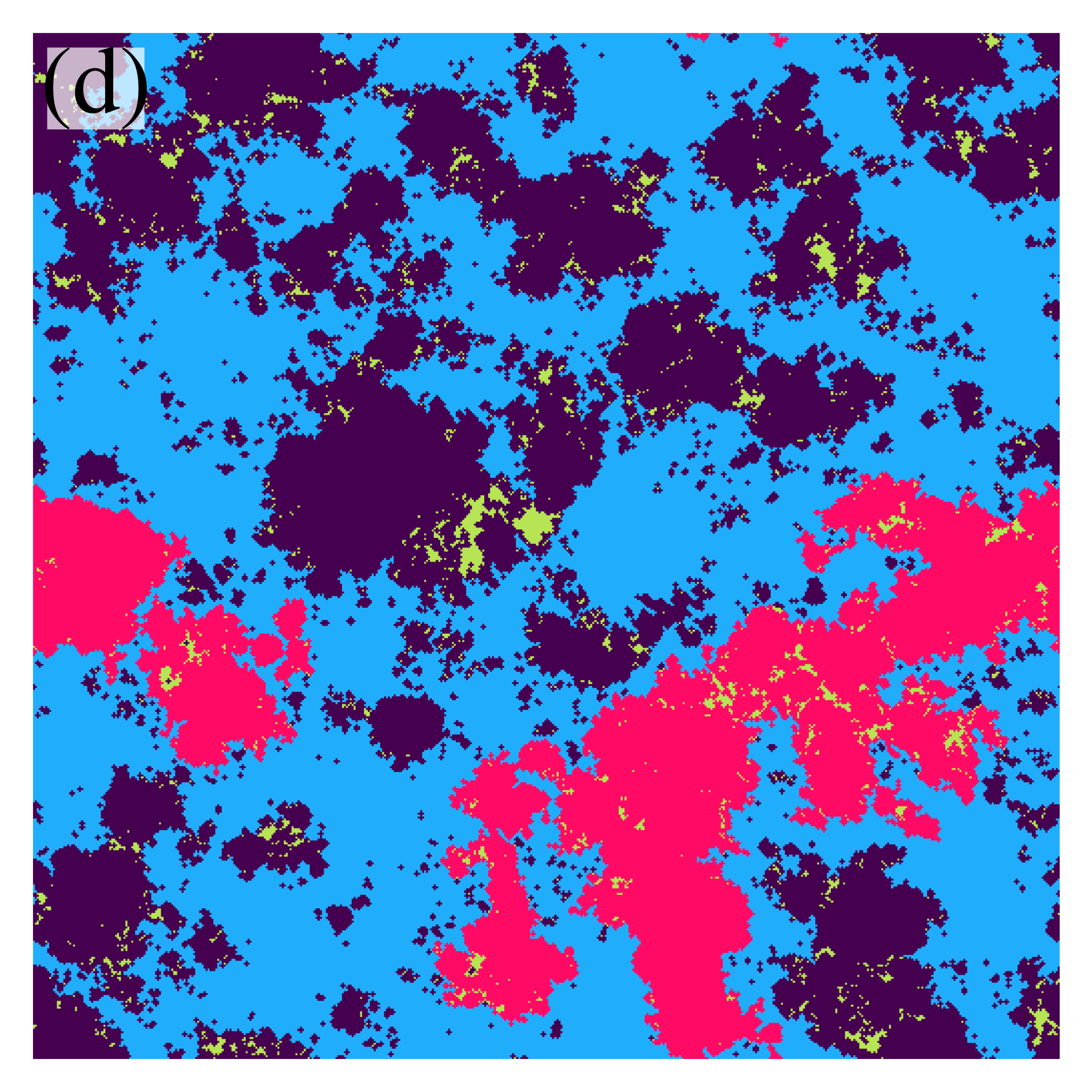}
    \includegraphics[width=0.157\textwidth]{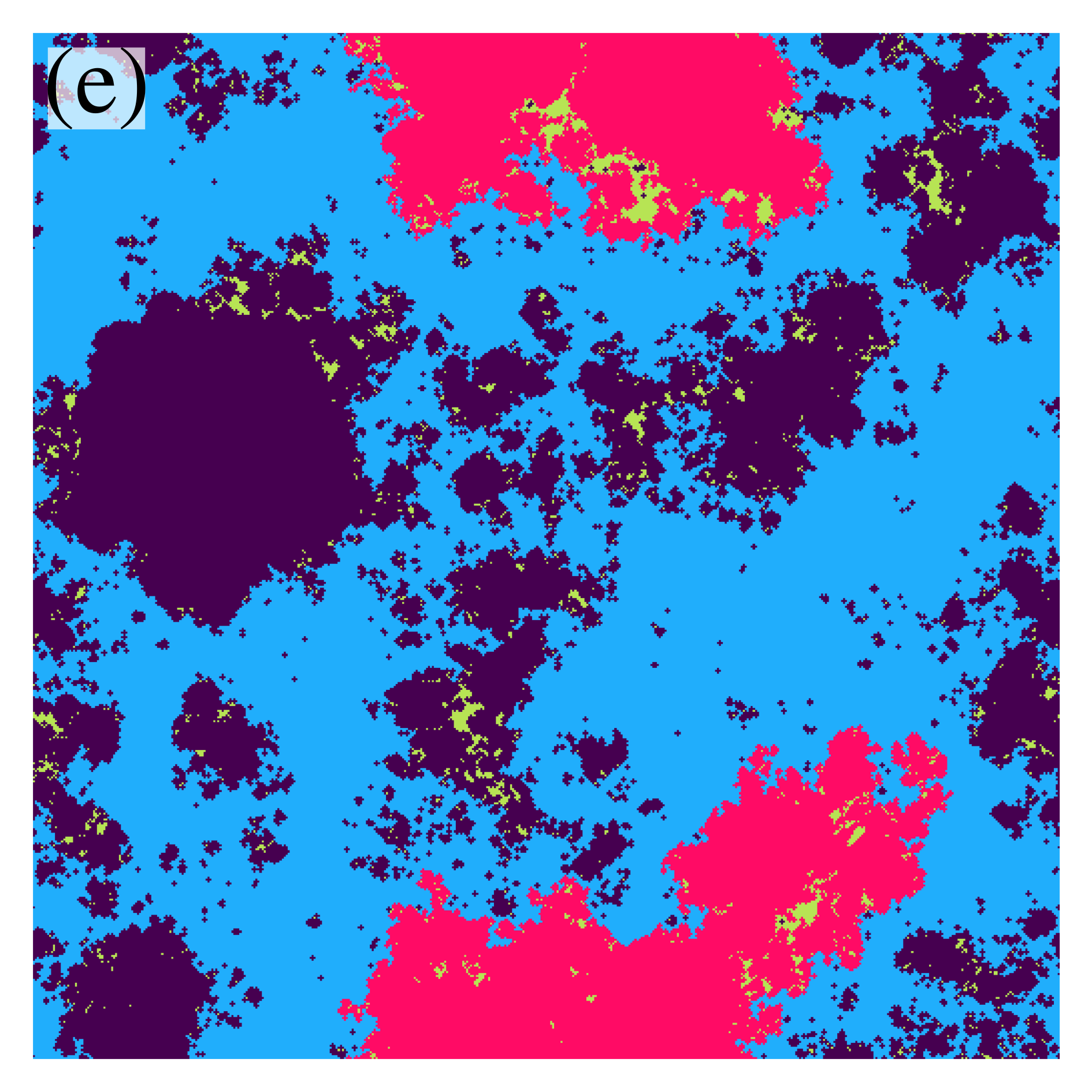}
    \includegraphics[width=0.157\textwidth]{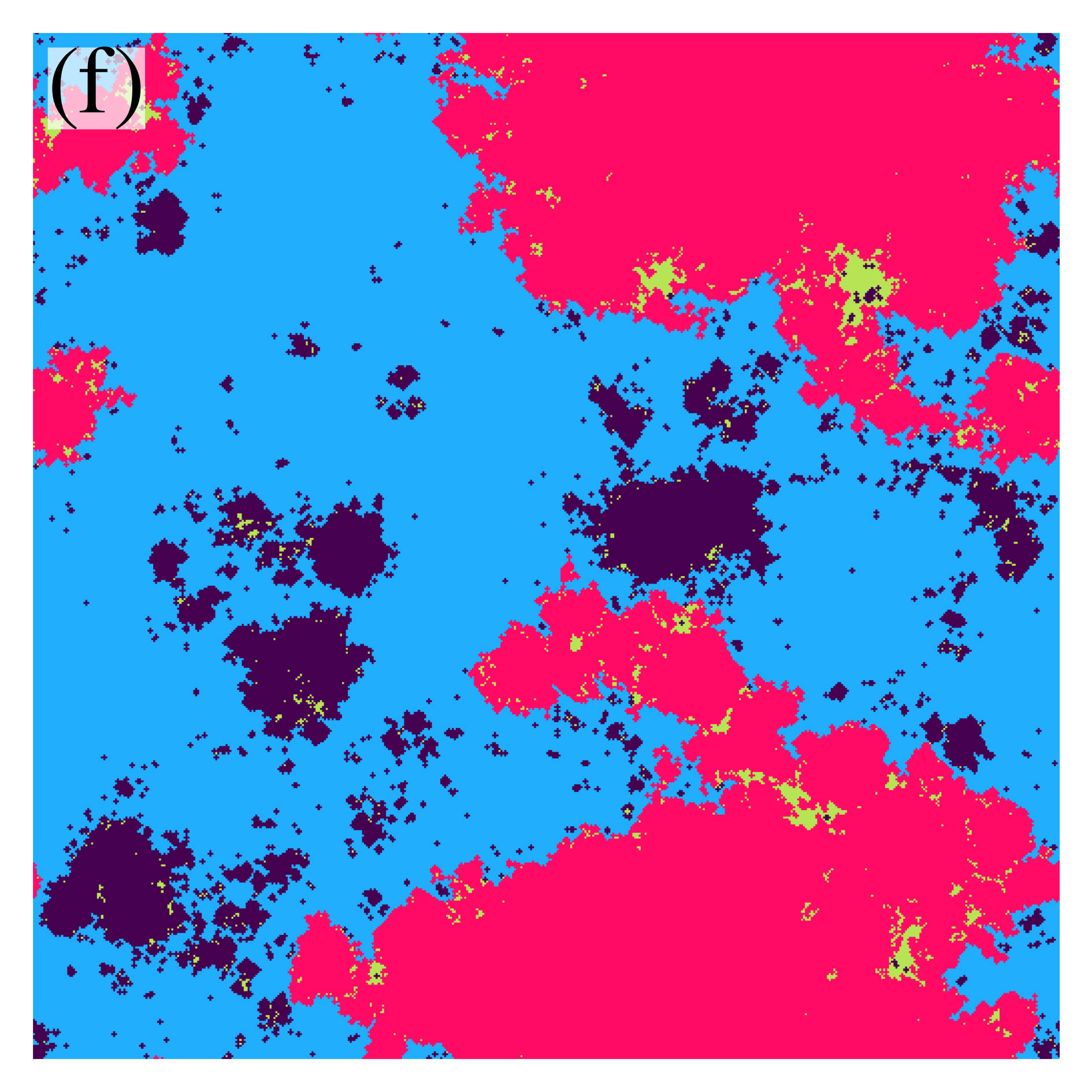}
    \caption{\label{fig_mosaic}
    Instantaneous configurations of the signs of the height fluctuations about their mean,
    during the kinetic roughening of a growing interface simulated by the SHE-Euler rule at 
    $t=$
    (a) 1, (b) 5, (c) 60, 
    (d) 480 (core of the growth regime), 
    (e) 3840 (transition to the steady state), 
    (f) 15360 (steady state).
    Positive (negative) signs are painted in violet (green) color.
    The largest cluster of positive (negative) sign is highlighted in magenta (blue) color.
    Periodic boundaries are employed.
    The lateral size is $L = 512$.
    }
    \end{figure}

Interfaces govern processes such as wetting, adhesion, protein adsorption, membrane transport, and phase separation. 
Understanding their structure and dynamics is essential for fields ranging from 
material science~\cite{Ohring_book02} and nanotechnology~\cite{Barth05},
to soft matter~\cite{Nagel17} and biophysics~\cite{Bru03}.
Growing interfaces have been intensively investigated due to the spontaneous emergence of scale invariance and their connection to other problems~\cite{HHealy95, Kriecherbauer10}.
Initially glimpsed in simple models~\cite{HHealy95}, 
universality in this field is now embodied by evidence from real complex systems~\cite{Maunuksela97, Takeuchi10, Ojeda00, Palasantzas02, Almeida14, HHealy14}.
%
%
In two-dimensional (2D) directed interfaces (that is, single-valued ones, with no overhangs nor crossings), 
clusters can be defined as the projection on the plane of connected regions lying either above or below a threshold~\cite{Kondev95}: 
Fig.~\ref{fig_mosaic} and Video 1 in the Supplementary Material (SM) provide examples thus constructed from a simulated, kinetically roughened growing interface.
%
%
The boundaries of the clusters have been studied for Gaussian interfaces at equilibrium~\cite{Kondev95, Olami96, Schwartz01, Schramm09, Giordanelli16}, 
and for growing interfaces in the steady state~\cite{Saberi08, Saberi09, Saberi10}.
%
%
However, their coarsening dynamics have not been studied yet.

In this work, we investigate the coarsening of clusters in 2D nonequilibrium interfaces growing by local rules. 
The aim is to characterize how domain coarsening emerges and whether it exhibits universal scaling behavior. 
We focus on standard lattice growth models that are widely believed to fall within the universality class of the Kardar–Parisi–Zhang (KPZ) equation \cite{Kardar86}, arguably the most influential continuum description of roughening and growth processes. 
Results for the simpler, linear KPZ counterpart, i.e., the Edwards-Wilkinson equation \cite{Edwards82}, can be found in the SM.

\textit{The Kardar-Parisi-Zhang universality class}.---
%
%
The KPZ equation \cite{Kardar86} 
is a stochastic model for the kinetic roughening \cite{Krug91}
of interfaces evolving by local aggregation. 
%
%
It is the prime object within the homonymous universality class, 
which accounts for the fluctuation properties of many macroscopic nonequilibrium 
or disordered systems \cite{HHealy95, Kriecherbauer10, Corwin11, HHealy15, *Takeuchi18},
both 
classical \cite{Krug91, HHealy95, Kriecherbauer10} and 
quantum \cite{Nahum17, 
Fontaine22, *Widmann25,
Ljubotina19, *Ye22, *Takeuchi25}.
%
%
The traditional picture is that of a $d$-dimensional interface of lateral size $L$ between a stable and an unstable phase.
%
%
A continuous field $h(\bm{r},t)$, 
measured normal to a reference profile, 
denotes the height of the configuration at position $\bm{r}$
and time $t$.
%
%
Apart from a constant growth velocity, which can be set to zero by using a coordinate system co-moving 
with the interface, the field evolves as~\cite{Kardar86}:
    \begin{equation}
    \partial_t h(\bm{r},t) = \nu \nabla^2 h + \frac{\lambda}{2}(\nabla h)^2 + \eta(\bm{r},t)\; ,
    \label{eq_KPZ}
    \end{equation}
    where the Laplacian comes from an elastic energy, 
    the quadratic slope term accounts for lateral growth,
    and the noise is Gaussian distributed with zero mean and $\overline{\eta(\bm{r},t)\eta(\bm{r}',t')} = 2D\delta^d(\bm{r}-\bm{r}')\delta(t-t')$.
    The overline denotes ensemble average; $\nu$, $\lambda$, and $D$ are parameters.
    For $\lambda = 0$, one recovers the Edwards-Wilkinson equation~\cite{Edwards82,Nattermann92}.
    %


The quadratic slope term breaks the up-down symmetry, $h \mapsto -h$, of Eq.~\eqref{eq_KPZ}.
%
%
%
%
From the flat initial profiles that we focus on, 
KPZ interfaces become statistically invariant under self-affine transformations,
$\bm{r}' \mapsto b\bm{r}$, $t' \mapsto b^zt$, $h'\mapsto b^{\alpha}h$ at long times~\cite{Krug91};
$b > 1$ is a scaling factor,
while $\alpha$ and $z$ are scaling exponents. 
%
%
By setting $b^zt \equiv 1$, then $\bm{r}' \mapsto \bm{r}/\xi(t)$,
where $\xi(t) \sim t^{1/z}$ is the correlation length. 
For finite systems, one distinguishes between 
the ``growth regime" $\xi \ll L$, 
and the steady state $\xi \sim L$.
Invariance of Eq.~(\ref{eq_KPZ}) under infinitesimal tilting leads to $\alpha + z = 2$~\cite{Kardar86}.
%
%
%
%
These exponents are not exactly known in 2D
\footnote{Some values predicted by 
mode-coupling~\cite{Doherty94, *Tu94, *Colaiori01}, 
operator product expansion~\cite{Lassig98}, 
nonperturbative renormalization group~\cite{Canet10, *Kloss12}, 
and conjectures~\cite{Wolf87, Kim89, Oliveira22}
are fairly close to recent numerical evaluations~\cite{Kelling11, *Kelling18, Pagnani15}.}.
%
%
Distributions of rescaled height fluctuations~\cite{Hhealy12, *HHealy13} 
and scaling forms of spatial correlators~\cite{HHealy14, Carrasco14, *Carrasco22, *Carrasco23} 
are numerically documented for several growth conditions.

%
Here, we uncover new universal features of 2D KPZ systems by analyzing the cluster dynamics in three growth models displaying KPZ scaling.
Clusters are defined through the sign of height fluctuations~\cite{Saberi08}, 
$\delta h(\bm{r},t) = h - \langle h \rangle$, 
relative to the mean interface height $\langle h \rangle$,
with $\langle \dots\rangle$ denoting spatial average.
%
%
%
Investigating the clusters' configurations, we describe the main aspects of the dynamics and seek for a manifestation of the up-down asymmetry of KPZ systems.
By evaluating a two-point spatial correlator, we probe the coarsening law and test the statistical time-invariance of the configurations.
As we shall see, the growth of the largest cluster in each population reveals a fundamental asymmetry.
We also characterize the populations of areas by their number densities.
Universal aspects are highlighted along the text. 
Below, we describe the methods.
Results and discussions are presented afterwards.


\textit{Methods}.---
%
The KPZ equation can be mapped onto the stochastic heat equation (SHE) with multiplicative noise. 
%
By discretizing the SHE 
with standard Euler schemes, 
transforming it back to the KPZ language, 
and setting $\lambda \rightarrow \infty$,
the two-stage growth rule is derived~\cite{Newman97}:
    \begin{equation}
    \begin{split}
    (i) \quad & \tilde{h}(\bm{r},t) = h(\bm{r},t) + \zeta(\bm{r},t)
    \; , \\
    (ii) \quad & h(\bm{r},t+1) = \max_{\bm{r}'\textrm{ nn }\bm{r}}[\tilde{h}(\bm{r},t), \tilde{h}(\bm{r}',t)]
    \; ,
    \end{split}
    \label{eq_dprm}
    \end{equation}
where $\zeta$ is as a random variable drawn from a Gaussian distribution with zero mean and unit variance.
In stage $(ii)$, the sites $\bm{r}'$ include all nearest neighbors of $\bm{r}$.
Unlike other discretization schemes \cite{Moser91}, the rule \eqref{eq_dprm} is free from instabilities.
We refer to it as the SHE-Euler model.

In the “restricted solid-on-solid" (RSOS) model \cite{Kim89}, 
a crystalline substrate is in contact with a supersaturated vapor. 
In an attempt of deposition, 
a randomly selected particle of unit volume from the gas follows a ballistic trajectory, 
normal to the substrate, 
until reaching the interface. 
The particle permanently attaches to the interface at the selected site $\bm{r}$, 
$h(\bm{r}) \rightarrow h(\bm{r}) + 1$, 
if the inequality $|h(\bm{r}) - h(\bm{r}')| \leq n$ holds after the possible deposition, 
$\forall \bm{r}'$ in the nearest neighborhood of $\bm{r}$; 
$n > 0$ is a restriction parameter. 
If the inequality is to be violated, the incoming particle is reflected back to the gas.

In a model for etched solids (ETC) \cite{Mello01}, a corrosive fluid wets the surface of a crystal.
The fluid advances by randomly dissolving a variable amount of unit volumes of the crystal in a microscopic timescale in which a site $\bm{r}$ is chosen at random;
the height $h(\bm{r}')$ of each $\bm{r}'$ in the nearest neighborhood of $\bm{r}$ is updated, 
$h(\bm{r}') \rightarrow h(\bm{r})$ if $h(\bm{r}') < h(\bm{r})$;
and the dissolution at the focal site is implemented, 
$h(\bm{r}) \rightarrow h(\bm{r}) + 1$.


The simulations were initialized with flat interfaces, 
$h(\bm{r},0)=0$, on a square lattice of size $L\times L$
and periodic boundaries.
%
The height variables in the SHE-Euler rule were updated in parallel,
while
the deposition attempts (corrosion rounds) in the RSOS (ETC) model 
were simulated sequentially.
$L^2$ sequences constitute a unit time step.
We set $n = 3$ 
in the RSOS because of a fast convergence to scaling \cite{Kim15}.
%
%
For all models, we verify that the interfacial width obeys dynamic scaling 
\cite{Family85} with the KPZ exponents [Fig.~S1, SM].
%
%
The number of samples used in ensemble averages are reported in the Sec. “Sampling" of the SM.
%
%
%
%
We attributed a binary variable $s(\bm{r},t)= \sgn[\delta h(\bm{r},t)]$ to $\delta h \neq 0$; 
otherwise, $s = \pm 1$ to $\delta h=0$ with the same probability.
A cluster is a connected region
of next neighboring sites with the same sign.
Clusters were detected by a labeling algorithm~\cite{Hoshen76}.

%
In the models in which $h \in \mathbb{N}$,
a large set of columns 
is inexorably above (below) $\langle h \rangle$
when the mean height is just below (above) an integer value.
As a consequence, 
in the clock that counts the attempts of deposition,
observables oscillate deterministically
because they are dictated by the fractional part of $\langle h \rangle$ [Fig.~S2, SM].
The appropriate handling is to monitor these systems using the clock of the laboratory, $t_{\textrm{lab}}(t) := \langle h \rangle \sim t$.
%
Ensemble averages then refer to sampling at the same $t_{\textrm{lab}}$. 
%
We use a precision of at least 4 digits after the decimal point of $t_{\textrm{lab}}$.
Measurements are performed at half-integer values of $t_{\textrm{lab}}$ for convenience since no 
$\delta h = 0$ event occurs.
With this procedure we recover the scaling laws measured in systems with real $h$.

In the following, we focus on the SHE-Euler rule.
%
Results for the other models are either included to support universality or 
can be found in the SM.


\textit{Mosaics}.---
%
Although the positive and negative signs initially appear to be well intermixed throughout the system, a difference in the size of the largest cluster in each population is already noticeable at early times [Fig.~\ref{fig_mosaic}(a)]. 
The asymmetry is enhanced in the subsequent evolution by a faster growth of one of these clusters, 
which quickly occupies nearly half of the system area and percolates the torus 
[blue cluster in Fig.~\ref{fig_mosaic}(b)].
On the contrary, 
its counterpart of opposite sign remains a non-percolating, small, and skinny cluster
during the same period [magenta cluster in Fig.~\ref{fig_mosaic}(b)].

The mosaics are noisy.
They contain tiny clusters [Fig.~\ref{fig_mosaic}(c)] 
that stochastically grow, shrink, and undergo ``birth'' and ``death'' processes [Video 1, SM].
%
%
Regardless of the cluster sizes, their boundaries are highly fluctuating.
Equal-sign clusters which are close by, often attach to each other [Video 1, SM].
Attachment may lead to the temporary formation of larger objects with a structure of blobs linked by thin bridges [e.g., the magenta cluster in Fig.~\ref{fig_mosaic}(d)].
Such structures are vigorously reshaped by the dynamics via attachment and detachment events, 
the reason why, from time to time, the largest cluster of positive sign is formed by dissimilar sets of sites [Figs.~\ref{fig_mosaic}(a)-(e); Video 2, SM].
%
%
The average size of positive-sign clusters increases 
up to the steady state (SS) [Fig.~\ref{fig_mosaic}].
The giant cluster, which is formed by negative signs in this case, also coarsens during most of the growth regime (GR)~---~as we shall quantify below~---~but the remaining area fraction of negative sign diminishes until the SS 
[compare the area occupied by the green clusters in Fig.~\ref{fig_mosaic}(b) and Fig.~\ref{fig_mosaic}(e)].
%
%
In the SS, the coarsening is interrupted. 
In this regime, few macroscopic clusters coexist punctured by a fluctuating set of smaller domains:
the giant percolating cluster formed at early times,
and counterparts of the opposite sign that slowly grew during the dynamics 
[there is only one counterpart, macroscopic cluster in the case shown in Fig.~\ref{fig_mosaic}(f)].
For the latter, the random attachment and detachment of nearby clusters remain active [Videos 3 and 4, SM].

Similar configurations are observed for the RSOS and ETC growth rules in the regime ($t_{\textrm{lab}} > 5$)
in which the density of sites of null height has vanished [Fig.~S3 and Fig.~S4, SM].
Visual inspection of the mosaics, 
along with quantitative results to be presented, 
point out that the giant cluster is harnessed to the sign of $\lambda$:
It is formed by negative (positive) height fluctuations
in the SHE-Euler and ETC (RSOS) models, 
for which $\lambda > 0$ ($\lambda < 0$).
No giant cluster is formed at early times if $\lambda = 0$ [Sec.~III.C, SM]. 

\textit{Two-point spatial correlator}.---
To quantify domain coarsening and test a possible statistical time-invariance of the mosaics, we analyze the following correlator:
    \begin{equation}
    C(r=|\bm{r}|,t) = \overline{\langle s(\bm{r}',t) s(\bm{r}'+\bm{r}, t) \rangle_{x,y} - \langle s \rangle^2}
    \; .
    \label{eq_correlation}
    \end{equation}
    For efficiency, we compute the sign-sign product only from the subset of pairs that lie on the same line ($x$ direction) or on the same column ($y$ direction).
    Isotropy is assumed.
    The symbol $\langle \dots \rangle_{x,y}$ indicates that the average is performed over such a subset.


    \begin{figure}[tb!]
    \includegraphics[width=0.23\textwidth]{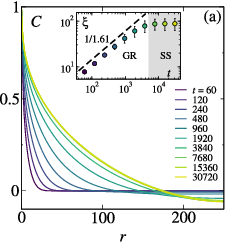}
    \includegraphics[width=0.232\textwidth]{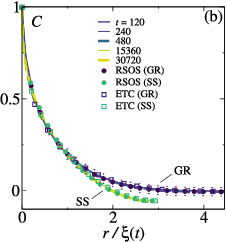}
    \caption{\label{fig_correlation}
    Space-time correlations  in the SHE-Euler model.
    (a) Data at different times given in the key, increasing from left to right.
    Inset: the correlation length versus time.
    The dashed line is a guide-to-the-eye with slope $1/z$ and $z \approx 1.61$.
    White and gray areas distinguish the growth regime (GR) and the steady state (SS), respectively.
    (b) Evidence for dynamic scaling and universality in the GR (blue data above) and SS (green data below).
    Lines refer to data from (a).
    Datapoints 
    for the RSOS  
    with $t_{\textrm{lab}} = 1600.5$ (GR), 51200.5 (SS);
    and ETC with $t_{\textrm{lab}} = 320.5$ (GR), $20480.5$ (SS).
    Uncertainties correspond to one standard deviation.
    %
    %
    }
    \end{figure}

The correlator decays monotonically from unity to a time-dependent value $C(L/2, t)$ [Fig.~\ref{fig_correlation}(a)].
%
%
By construction, $C(L/2, t \ll t^*) = 0$, where $t^* \sim L^z$.
In the fully correlated region, 
$C(L/2, t \gg t^*) \approx -0.06$ for $L \geq 32$.
In the transition from one regime to another, 
the large-distance correlator interpolates between the two foregoing outcomes [$960 \lesssim t \lesssim 3840$ in Fig.~\ref{fig_correlation}(a)].
Pairs of signs become slightly anti-correlated for $r \sim L/2$ because of the coexistence of few macroscopic clusters of opposite sign in the SS.

On the one hand, the decay of $C(r,t \ll t^*)$ slows down over time owing to coarsening.
On the other hand, a time-independent form is achieved in the SS.
By defining $C(\xi, t)\equiv 0.2$, 
as usual in many nonequilibrium systems \cite{Corberi11}, 
we verify the growth law $\xi(t) \sim t^{1/z}$ with $z \approx 1.61$~\cite{Pagnani15}, 
and the saturation level $\xi(t) = \textrm{constant}$ [inset, Fig.~\ref{fig_correlation}(a)].
In the core of both regimes, 
the correlator is statistically time-invariant under the normalized scale $r/\xi(t)$ [Fig.~\ref{fig_correlation}(b)].
The master curves for $C(r/\xi)$ are distinct in the growth and saturation regimes.
Within our numerical precision, the corresponding scaling functions for the SHE-Euler growth rule are identical to those obtained from the RSOS and ETC models 
[Fig.~\ref{fig_correlation}(b)], 
thus signaling universality. 
These two universal forms are distinct from the ones that characterize 
the two-point spatial correlator of the height fluctuations [Fig.~S5, SM].

\begin{figure}[bt!]
    \includegraphics[width=0.23\textwidth]{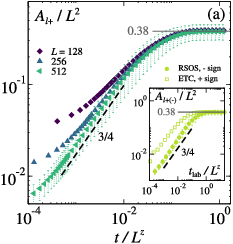}
    \includegraphics[width=0.23\textwidth]{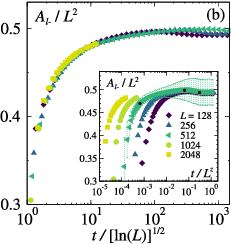}
    \caption{\label{fig_largestarea}
    Ensemble-averaged area of the largest cluster of positive (a) and negative (b) sign in the SHE-Euler model. 
    Inset of (a): $A_{l-(+)}$ area of the largest cluster of negative (positive) sign in the RSOS (ETC) model for $L = 512$.
    Solid lines indicate the asymptotic area fraction, common to all models.
    Dashed lines are guides-to-the-eye with the slopes depicted.
    Insertion in (b): same data of the main panel, scaled so as to collapse data at long times.
    Black dots are guides-to-the-eye that distinguish among segments of the curve.
    Dotted lines cover the interval of the representative uncertainty, which is defined as one standard deviation.
    }
    \end{figure}

\textit{Area occupied by the largest clusters}.---%
We monitored the ensemble-averaged area of the largest cluster of positive (negative) sign, $A_{l+(-)}(L,t)$.
%
%
The area fraction $A_{l+}/L^2$, 
observed through the scale $t/L^z$, 
increases monotonically towards 
its SS limit $c = 0.38 \pm 0.07$ [Fig.~\ref{fig_largestarea}(a)].
This is the same value reached by the corresponding area fractions of the RSOS and ETC models [inset of Fig.~\ref{fig_largestarea}(a)].
(In the RSOS, the populations are inverted since $\lambda < 0$.) 
Finite-size effects are present at early times only;
the smaller the $L$, the higher the area fraction.
Data collapse on a master curve at sufficiently late times.
In the core of the growth regime, 
$A_{l+}(L,t) \sim t^{\gamma} L^{2-\gamma z}$ 
with $\gamma \approx 3/4$ [Fig.~\ref{fig_largestarea}(a)].
An exponent $\gamma \approx 3/4$ describes the RSOS and ETC data  as well [inset, Fig.~\ref{fig_largestarea}(a)].
In summary, 
$A_{l+}(L,t) \simeq c L^2 \mathcal{A}_{+}[\theta = t/(c'L^z)]$,
where $c'$ is a model-dependent parameter,
$\mathcal{A}_{+}(\theta \ll 1) \sim \theta^{\gamma}$, and
$\mathcal{A}_{+}(\theta \gg 1) = 1$.

The area fraction $A_{l-}/L^2$
has a non-monotonic behavior which can be segmented into four characteristic stages [inset of Fig.~\ref{fig_largestarea}(b)].
Firstly, it grows unusually fast up to $\approx0.47$. 
From there, a slower growth takes place until it stops at the maximum, $\approx 0.50$.
A slight decrease towards the SS outcome, $\approx 0.488-.499$ (idem for RSOS and ETC), 
then precedes the plateau.
Although the two final parts are collapsed under the scale $t/L^z$, the initial regimes are not [inset of Fig.~\ref{fig_largestarea}(b)].
Such initial regimes become independent of $L$ through the ratio $t/\sqrt{\ln L}$ [Fig.~\ref{fig_largestarea}(b)].
The same ratio collapses the corresponding data in the RSOS and ETC models 
[Fig.~S6, SM].
The novel scaling implies the existence of a growing length 
$\xi_{\textrm{g}}(t) \sim e^{(t/t_a)^2}$ associated to 
the size of the giant cluster, where 
$t_a$ is a microscopic time. 
If $\lambda = 0$, no exponential growth is observed [Fig.~S10(a), SM].
Such growth law is proper of KPZ systems
and contrasts with the power-law 
$\sim t^{1/z_{\textrm{p}}}$
found in standard coarsening systems, 
where $z_{\textrm{p}} < z$ is a 
lattice-dependent exponent~\cite{Blanchard14, Tartaglia15, Blanchard17, Tartaglia18}. See also \cite{Blanchard12, Almeida25}.


\begin{figure}[bt!]
    \includegraphics[width=0.23\textwidth]{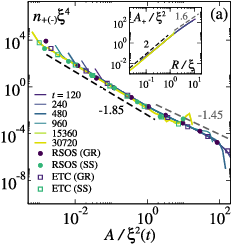}
    \includegraphics[width=0.23\textwidth]{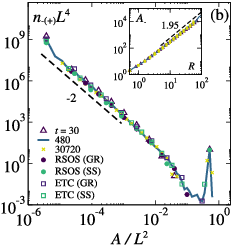}
    \caption{\label{fig_ndensity}
    Scaling forms for the number density of areas from (a) positive and (b) negative signs in the SHE-Euler model.
    Corresponding outcomes from RSOS [($t_{\textrm{lab}} = 1600.5$; 51200.5), (a) negative (b) positive sign] 
    and the ETC [($t_{\textrm{lab}} = 320.5$; $20480.5$), (a) positive (b) negative sign] 
    are shown.
    Insets show the area-radius relation computed from clusters in the population of positive (a) and negative (b) signs in the SHE-Euler model.
    The convention is the one of the keys in the main panels.
    Dashed lines are guides-to-the-eyes with the indicated slopes.
    }
    \end{figure}

\textit{Number density}.---
To quantify the population of areas, 
we count the number of clusters of positive (negative) sign per system area at $t$ with area between $A$ and $A + dA$, $n_{+(-)}(A,t)dA$.
%
%
For the population with no giant cluster, 
dynamic scaling suggests that areas should be measured in terms of $\xi^2(t)$.
Therefore, we analyze the dimensionless quantity $F = n_{+}(A,t)\xi^4(t)$ versus $u = A/\xi^2(t)$ [Fig.~\ref{fig_ndensity}(a)].
In this form, 
curves are collapsed on a $F_{\textrm{GR}}(u) \sim u^{-\tau_{1}}$ for $u \ll 1$;
$\sim u^{-\tau_{2}}$ for $u \gg 1$, 
where $\tau_1 \approx 1.85$ and $\tau_2 \approx 1.45$~\footnote{Although the data are compatible with an algebraic regime ruled by $\tau_2$, it is not fully resolved yet whether, due to finite-size effects, this is an artifact of a regime preceding the exponential decay.}.
There is also an exponentially fast decay after the second regime.
Such a decay gives rise to a peak for $t \gtrsim 960$ (a model- and size-dependent time).
Due to its location, the peak signals the presence of a recurrent macroscopic cluster in the mosaic.
In the SS, 
$F_{\textrm{SS}}(u) = F_{\textrm{GR}}(u)$ for $u \ll 1$, while the functions become distinguishable from $u \sim 1$ onward.
%
%
Evidence for the universality of
$F_{\textrm{GR}}$ and $F_{\textrm{SS}}$
comes from the fact that RSOS and ETC results are well described by the same functions
[Fig.~\ref{fig_ndensity}(a)].
Data collapse is observed over 10 decades in $F$ and nearly 5 decades in $u$.

In the population of the giant cluster, 
the key scale is not $\xi^2(t)$, but $\xi^2_g(t)$.
Since $\xi^2_g(t)$ soon reaches $L^2/2$, 
it is useful to gaze at $G = n_-(A,t)L^4$ versus $v = A/L^2$ [Fig.~\ref{fig_ndensity}(b)].
We checked that 
$(i)$ $G(v)$ is indeed independent of $L$ (not shown for the sake of visibility);
$(ii)$ a rescaling using $\xi(t)$ instead of $\xi_g(t)$ does not lead to data collapse.
Data at $t = 30$ 
are slightly above the solid line in the region $v \ll 10^{-2}$ because the population of small clusters  
diminishes at early times.
Apart from that, $G(v) \sim v^{-\tau_3}$, 
with $\tau_3 \approx 2$, 
for $v \ll 2 \times 10^{-3}$; 
after, there is a faster decay that precedes the peak at $v \sim 1/2$.
$G$ has the same shape in both the core of the GR and in the SS.
Its shape fits well the corresponding data from the RSOS and ETC models [Fig.~\ref{fig_ndensity}(b)].

For self-affine interfaces at equilibrium, $n(R)dR \sim R^{-2 +\alpha}$~\cite{Kondev95}, where $R \gg 1$ is the average radius of a cluster.
Since there is a one-to-one relation between radius and area, $n(A)dA = n(R)dR$.
Writing the area-radius relation as $A \sim R^{d_{\textrm{c}}}$, 
where $d_\textrm{c}$ is the dimension of the clusters, then $n(A) \sim A^{(-2+\alpha-d_{\textrm{c}})/d_{\textrm{c}}}$.
Taking into account a nonequilibrium problem with a growing correlation length, we write
$A/\xi^2(t) \sim (R/\xi)^{d_{\textrm{c}}}$,
and 
$n(A,t)\xi^4(t) \sim [A/\xi^2(t)]^{(-2+\alpha-d_{\textrm{c}})/d_{\textrm{c}}}$ for $A \ll \xi^2$.
%
%
%
%
%
For the KPZ case, we shall distinguish between the populations of signs.
For the population dictated by $\xi$, $d_{\textrm{c}} = 2$ for $R/\xi \ll 1$,
but $d_{\textrm{c}} = 1.6$ for $R/\xi \gg 1$ [inset of Fig.~\ref{fig_ndensity}(a)].
In the simulations, we regard $R$ as the radius of gyration of a cluster [Sec.~III.~F, SM].
Using the exact KPZ relation $\alpha + z = 2$, 
one has $\tau_1 = (z + 2)/2 \approx 1.81$~\footnote{Remarkably, this value is very close to the exponent of the distribution of holes within the critical percolation backbone, $\tau_b\simeq 1.82$~\cite{HuZiDe16}.}.
This value is close to our numerical estimate, and distinct from the (possibly) exact EW outcome $31/15 = 2.06...$ \cite{Saberi08, Duplantier90} [Fig.~S10(b), SM].
A similar argument cannot account for $\tau_2$ because it characterizes scales larger than $\xi^2$. 
The scaling argument does not hold for the atypical population of the giant cluster either: $d_{\textrm{c}} \approx 2$ [inset of Fig.~\ref{fig_ndensity}(b)], but $\tau_3 \neq \tau_1$.


\textit{Conclusions}.---%
We have studied the coarsening of clusters in 2D nonequilibrium interfaces growing with local rules.
Interfaces were simulated using lattice growth models widely believed to be in the KPZ universality class.
%
%
The regions of fluctuations about the surface average with 
sign contrary to the one of $\lambda$---the nonlinearity strength in the KPZ equation---form a giant cluster that quickly occupies nearly half of the system area.
On the other hand, 
clusters of sign equal to $\sgn(\lambda)$ have a slower growth,
albeit they also coarsen and coalesce with others with the same sign.
%
%
A two-point correlator demonstrates the statistical time-invariance of the sign configurations; 
their scaling forms are common to all models.
%
%
The growth of the largest cluster is sign-dependent: 
While it is dictated by the correlation length for the cluster of sign equal to $\sgn(\lambda)$, 
it grows exponentially fast for the giant cluster.
Such exponential law is proper of KPZ systems, since it is absent for $\lambda = 0$ [Fig.~S10, SM], and indeed unusual in coarsening systems.
%
%
Number densities of areas have sign-dependent scaling forms that are shared among different models.
These forms are distinct from the symmetric ones found
in the Edwards-Wilkinson dynamics [Fig.~S10(b), SM].
%
Our results identify new universal aspects of growing interfaces.
%
%
They invite experimental tests.

\begin{acknowledgments}
Research funded by the ManyBodyNet ANR-24-CE30-5851 grant (LFC), 
the Brazilian National Council for Scientific and Technological Development – CNPq Grant No.~443517/2023-1 (RALA, JJA, LFC)
and the Coordenação de Aperfeiçoamento de Pessoal de Nível Superior - Brasil (CAPES) - Finance Code 001 within the CAPES/COFECUB project 88887.987311/2024-00
(JJA, LFC), and  
CNPq, Fapemig (TJO).
We thank M. A. Moore, M. Picco and G. Schehr for very useful discussions.
\end{acknowledgments}

\bibliography{refs}

\providecommand{\noopsort}[1]{}\providecommand{\singleletter}[1]{#1}%
\begin{thebibliography}{81}%
\makeatletter
\providecommand \@ifxundefined [1]{%
 \@ifx{#1\undefined}
}%
\providecommand \@ifnum [1]{%
 \ifnum #1\expandafter \@firstoftwo
 \else \expandafter \@secondoftwo
 \fi
}%
\providecommand \@ifx [1]{%
 \ifx #1\expandafter \@firstoftwo
 \else \expandafter \@secondoftwo
 \fi
}%
\providecommand \natexlab [1]{#1}%
\providecommand \enquote  [1]{``#1''}%
\providecommand \bibnamefont  [1]{#1}%
\providecommand \bibfnamefont [1]{#1}%
\providecommand \citenamefont [1]{#1}%
\providecommand \href@noop [0]{\@secondoftwo}%
\providecommand \href [0]{\begingroup \@sanitize@url \@href}%
\providecommand \@href[1]{\@@startlink{#1}\@@href}%
\providecommand \@@href[1]{\endgroup#1\@@endlink}%
\providecommand \@sanitize@url [0]{\catcode `\\12\catcode `\$12\catcode `\&12\catcode `\#12\catcode `\^12\catcode `\_12\catcode `\%12\relax}%
\providecommand \@@startlink[1]{}%
\providecommand \@@endlink[0]{}%
\providecommand \url  [0]{\begingroup\@sanitize@url \@url }%
\providecommand \@url [1]{\endgroup\@href {#1}{\urlprefix }}%
\providecommand \urlprefix  [0]{URL }%
\providecommand \Eprint [0]{\href }%
\providecommand \doibase [0]{https://doi.org/}%
\providecommand \selectlanguage [0]{\@gobble}%
\providecommand \bibinfo  [0]{\@secondoftwo}%
\providecommand \bibfield  [0]{\@secondoftwo}%
\providecommand \translation [1]{[#1]}%
\providecommand \BibitemOpen [0]{}%
\providecommand \bibitemStop [0]{}%
\providecommand \bibitemNoStop [0]{.\EOS\space}%
\providecommand \EOS [0]{\spacefactor3000\relax}%
\providecommand \BibitemShut  [1]{\csname bibitem#1\endcsname}%
\let\auto@bib@innerbib\@empty
\bibitem [{\citenamefont {Goldenfeld}(2018)}]{Goldenfeld_book18}%
  \BibitemOpen
  \bibfield  {author} {\bibinfo {author} {\bibfnamefont {N.}~\bibnamefont {Goldenfeld}},\ }\href@noop {} {\emph {\bibinfo {title} {Lectures On Phase Transitions And The Renormalization Group}}}\ (\bibinfo  {publisher} {CRC Press},\ \bibinfo {year} {2018})\BibitemShut {NoStop}%
\bibitem [{\citenamefont {Bray}(1994)}]{Bray94}%
  \BibitemOpen
  \bibfield  {author} {\bibinfo {author} {\bibfnamefont {A.}~\bibnamefont {Bray}},\ }\bibfield  {title} {\bibinfo {title} {Theory of phase-ordering kinetics},\ }\href {https://doi.org/10.1080/00018739400101505} {\bibfield  {journal} {\bibinfo  {journal} {Adv. Phys.}\ }\textbf {\bibinfo {volume} {43}},\ \bibinfo {pages} {357} (\bibinfo {year} {1994})}\BibitemShut {NoStop}%
\bibitem [{\citenamefont {Halpin-Healy}\ and\ \citenamefont {Zhang}(1995)}]{HHealy95}%
  \BibitemOpen
  \bibfield  {author} {\bibinfo {author} {\bibfnamefont {T.}~\bibnamefont {Halpin-Healy}}\ and\ \bibinfo {author} {\bibfnamefont {Y.-C.}\ \bibnamefont {Zhang}},\ }\bibfield  {title} {\bibinfo {title} {{Kinetic roughening phenomena, stochastic growth, directed polymers and all that. Aspects of multidisciplinary statistical mechanics}},\ }\href {https://doi.org/https://doi.org/10.1016/0370-1573(94)00087-J} {\bibfield  {journal} {\bibinfo  {journal} {Phys. Rep.}\ }\textbf {\bibinfo {volume} {254}},\ \bibinfo {pages} {215} (\bibinfo {year} {1995})}\BibitemShut {NoStop}%
\bibitem [{\citenamefont {Henkel}\ \emph {et~al.}(2008)\citenamefont {Henkel}, \citenamefont {Hinrichsen},\ and\ \citenamefont {L{\"u}beck}}]{Henkel_book08}%
  \BibitemOpen
  \bibfield  {author} {\bibinfo {author} {\bibfnamefont {M.}~\bibnamefont {Henkel}}, \bibinfo {author} {\bibfnamefont {H.}~\bibnamefont {Hinrichsen}},\ and\ \bibinfo {author} {\bibfnamefont {S.}~\bibnamefont {L{\"u}beck}},\ }\href@noop {} {\emph {\bibinfo {title} {Non-Equilibrium Phase Transitions: Volume I: Absorbing Phase Transitions}}}\ (\bibinfo  {publisher} {Springer},\ \bibinfo {year} {2008})\BibitemShut {NoStop}%
\bibitem [{\citenamefont {Henkel}\ and\ \citenamefont {Pleimling}(2011)}]{Henkel_book11}%
  \BibitemOpen
  \bibfield  {author} {\bibinfo {author} {\bibfnamefont {M.}~\bibnamefont {Henkel}}\ and\ \bibinfo {author} {\bibfnamefont {M.}~\bibnamefont {Pleimling}},\ }\href@noop {} {\emph {\bibinfo {title} {Non-Equilibrium Phase Transitions: Volume 2: Ageing and Dynamical Scaling Far from Equilibrium}}}\ (\bibinfo  {publisher} {Springer Science \& Business Media},\ \bibinfo {year} {2011})\BibitemShut {NoStop}%
\bibitem [{\citenamefont {Arenzon}\ \emph {et~al.}(2007)\citenamefont {Arenzon}, \citenamefont {Bray}, \citenamefont {Cugliandolo},\ and\ \citenamefont {Sicilia}}]{Arenzon07}%
  \BibitemOpen
  \bibfield  {author} {\bibinfo {author} {\bibfnamefont {J.~J.}\ \bibnamefont {Arenzon}}, \bibinfo {author} {\bibfnamefont {A.~J.}\ \bibnamefont {Bray}}, \bibinfo {author} {\bibfnamefont {L.~F.}\ \bibnamefont {Cugliandolo}},\ and\ \bibinfo {author} {\bibfnamefont {A.}~\bibnamefont {Sicilia}},\ }\bibfield  {title} {\bibinfo {title} {{Exact Results for Curvature-Driven Coarsening in Two Dimensions}},\ }\href {https://doi.org/10.1103/PhysRevLett.98.145701} {\bibfield  {journal} {\bibinfo  {journal} {Phys. Rev. Lett.}\ }\textbf {\bibinfo {volume} {98}},\ \bibinfo {pages} {145701} (\bibinfo {year} {2007})}\BibitemShut {NoStop}%
\bibitem [{\citenamefont {Dornic}\ \emph {et~al.}(2001)\citenamefont {Dornic}, \citenamefont {Chat\'e}, \citenamefont {Chave},\ and\ \citenamefont {Hinrichsen}}]{Dornic01}%
  \BibitemOpen
  \bibfield  {author} {\bibinfo {author} {\bibfnamefont {I.}~\bibnamefont {Dornic}}, \bibinfo {author} {\bibfnamefont {H.}~\bibnamefont {Chat\'e}}, \bibinfo {author} {\bibfnamefont {J.}~\bibnamefont {Chave}},\ and\ \bibinfo {author} {\bibfnamefont {H.}~\bibnamefont {Hinrichsen}},\ }\bibfield  {title} {\bibinfo {title} {Critical coarsening without surface tension: The universality class of the voter model},\ }\href {https://doi.org/10.1103/PhysRevLett.87.045701} {\bibfield  {journal} {\bibinfo  {journal} {Phys. Rev. Lett.}\ }\textbf {\bibinfo {volume} {87}},\ \bibinfo {pages} {045701} (\bibinfo {year} {2001})}\BibitemShut {NoStop}%
\bibitem [{\citenamefont {Sicilia}\ \emph {et~al.}(2008)\citenamefont {Sicilia}, \citenamefont {Arenzon}, \citenamefont {Dierking}, \citenamefont {Bray}, \citenamefont {Cugliandolo}, \citenamefont {Mart\'{\i}nez-Perdiguero}, \citenamefont {Alonso},\ and\ \citenamefont {Pintre}}]{Sicilia08}%
  \BibitemOpen
  \bibfield  {author} {\bibinfo {author} {\bibfnamefont {A.}~\bibnamefont {Sicilia}}, \bibinfo {author} {\bibfnamefont {J.~J.}\ \bibnamefont {Arenzon}}, \bibinfo {author} {\bibfnamefont {I.}~\bibnamefont {Dierking}}, \bibinfo {author} {\bibfnamefont {A.~J.}\ \bibnamefont {Bray}}, \bibinfo {author} {\bibfnamefont {L.~F.}\ \bibnamefont {Cugliandolo}}, \bibinfo {author} {\bibfnamefont {J.}~\bibnamefont {Mart\'{\i}nez-Perdiguero}}, \bibinfo {author} {\bibfnamefont {I.}~\bibnamefont {Alonso}},\ and\ \bibinfo {author} {\bibfnamefont {I.~C.}\ \bibnamefont {Pintre}},\ }\bibfield  {title} {\bibinfo {title} {Experimental test of curvature-driven dynamics in the phase ordering of a two dimensional liquid crystal},\ }\href {https://doi.org/10.1103/PhysRevLett.101.197801} {\bibfield  {journal} {\bibinfo  {journal} {Phys. Rev. Lett.}\ }\textbf {\bibinfo {volume} {101}},\ \bibinfo {pages} {197801} (\bibinfo {year} {2008})}\BibitemShut {NoStop}%
\bibitem [{\citenamefont {Almeida}(2023)}]{Almeida23}%
  \BibitemOpen
  \bibfield  {author} {\bibinfo {author} {\bibfnamefont {R.~A.~L.}\ \bibnamefont {Almeida}},\ }\bibfield  {title} {\bibinfo {title} {Critical percolation in the ordering kinetics of twisted nematic phases},\ }\href {https://doi.org/10.1103/PhysRevLett.131.268101} {\bibfield  {journal} {\bibinfo  {journal} {Phys. Rev. Lett.}\ }\textbf {\bibinfo {volume} {131}},\ \bibinfo {pages} {268101} (\bibinfo {year} {2023})}\BibitemShut {NoStop}%
\bibitem [{\citenamefont {Almeida}\ and\ \citenamefont {Arenzon}(2025)}]{Almeida25}%
  \BibitemOpen
  \bibfield  {author} {\bibinfo {author} {\bibfnamefont {R.~A.~L.}\ \bibnamefont {Almeida}}\ and\ \bibinfo {author} {\bibfnamefont {J.~J.}\ \bibnamefont {Arenzon}},\ }\bibfield  {title} {\bibinfo {title} {Critical clusters in liquid crystals: Fractal geometry and conformal invariance},\ }\href {https://doi.org/10.1103/PhysRevLett.134.178101} {\bibfield  {journal} {\bibinfo  {journal} {Phys. Rev. Lett.}\ }\textbf {\bibinfo {volume} {134}},\ \bibinfo {pages} {178101} (\bibinfo {year} {2025})}\BibitemShut {NoStop}%
\bibitem [{\citenamefont {Prozorov}\ \emph {et~al.}(2008)\citenamefont {Prozorov}, \citenamefont {Fidler}, \citenamefont {Hoberg},\ and\ \citenamefont {Canfield}}]{Prozorov08}%
  \BibitemOpen
  \bibfield  {author} {\bibinfo {author} {\bibfnamefont {R.}~\bibnamefont {Prozorov}}, \bibinfo {author} {\bibfnamefont {A.~F.}\ \bibnamefont {Fidler}}, \bibinfo {author} {\bibfnamefont {J.~R.}\ \bibnamefont {Hoberg}},\ and\ \bibinfo {author} {\bibfnamefont {P.~C.}\ \bibnamefont {Canfield}},\ }\bibfield  {title} {\bibinfo {title} {Suprafroth in type-{I} superconductors},\ }\href {https://doi.org/10.1038/nphys888} {\bibfield  {journal} {\bibinfo  {journal} {Nature Phys.}\ }\textbf {\bibinfo {volume} {4}},\ \bibinfo {pages} {327} (\bibinfo {year} {2008})}\BibitemShut {NoStop}%
\bibitem [{\citenamefont {Lambert}\ \emph {et~al.}(2010)\citenamefont {Lambert}, \citenamefont {Mokso}, \citenamefont {Cantat}, \citenamefont {Cloetens}, \citenamefont {Glazier}, \citenamefont {Graner},\ and\ \citenamefont {Delannay}}]{Lambert10}%
  \BibitemOpen
  \bibfield  {author} {\bibinfo {author} {\bibfnamefont {J.}~\bibnamefont {Lambert}}, \bibinfo {author} {\bibfnamefont {R.}~\bibnamefont {Mokso}}, \bibinfo {author} {\bibfnamefont {I.}~\bibnamefont {Cantat}}, \bibinfo {author} {\bibfnamefont {P.}~\bibnamefont {Cloetens}}, \bibinfo {author} {\bibfnamefont {J.~A.}\ \bibnamefont {Glazier}}, \bibinfo {author} {\bibfnamefont {F.}~\bibnamefont {Graner}},\ and\ \bibinfo {author} {\bibfnamefont {R.}~\bibnamefont {Delannay}},\ }\bibfield  {title} {\bibinfo {title} {Coarsening foams robustly reach a self-similar growth regime},\ }\href {https://doi.org/10.1103/PhysRevLett.104.248304} {\bibfield  {journal} {\bibinfo  {journal} {Phys. Rev. Lett.}\ }\textbf {\bibinfo {volume} {104}},\ \bibinfo {pages} {248304} (\bibinfo {year} {2010})}\BibitemShut {NoStop}%
\bibitem [{\citenamefont {McNally}\ \emph {et~al.}(2017)\citenamefont {McNally}, \citenamefont {Bernardy}, \citenamefont {Thomas}, \citenamefont {Kalziqi}, \citenamefont {Pentz}, \citenamefont {Brown}, \citenamefont {Hammer}, \citenamefont {Yunker},\ and\ \citenamefont {Ratcliff}}]{Mcnally17}%
  \BibitemOpen
  \bibfield  {author} {\bibinfo {author} {\bibfnamefont {L.}~\bibnamefont {McNally}}, \bibinfo {author} {\bibfnamefont {E.}~\bibnamefont {Bernardy}}, \bibinfo {author} {\bibfnamefont {J.}~\bibnamefont {Thomas}}, \bibinfo {author} {\bibfnamefont {A.}~\bibnamefont {Kalziqi}}, \bibinfo {author} {\bibfnamefont {J.}~\bibnamefont {Pentz}}, \bibinfo {author} {\bibfnamefont {S.~P.}\ \bibnamefont {Brown}}, \bibinfo {author} {\bibfnamefont {B.~K.}\ \bibnamefont {Hammer}}, \bibinfo {author} {\bibfnamefont {P.~J.}\ \bibnamefont {Yunker}},\ and\ \bibinfo {author} {\bibfnamefont {W.~C.}\ \bibnamefont {Ratcliff}},\ }\bibfield  {title} {\bibinfo {title} {{Killing by Type {VI} secretion drives genetic phase separation and correlates with increased cooperation}},\ }\href {https://doi.org/10.1038/ncomms14371} {\bibfield  {journal} {\bibinfo  {journal} {Nature Comm.}\ }\textbf {\bibinfo {volume} {8}},\ \bibinfo {pages} {14371} (\bibinfo {year} {2017})}\BibitemShut {NoStop}%
\bibitem [{\citenamefont {Sicilia}\ \emph {et~al.}(2007)\citenamefont {Sicilia}, \citenamefont {Arenzon}, \citenamefont {Bray},\ and\ \citenamefont {Cugliandolo}}]{Sicilia07}%
  \BibitemOpen
  \bibfield  {author} {\bibinfo {author} {\bibfnamefont {A.}~\bibnamefont {Sicilia}}, \bibinfo {author} {\bibfnamefont {J.~J.}\ \bibnamefont {Arenzon}}, \bibinfo {author} {\bibfnamefont {A.~J.}\ \bibnamefont {Bray}},\ and\ \bibinfo {author} {\bibfnamefont {L.~F.}\ \bibnamefont {Cugliandolo}},\ }\bibfield  {title} {\bibinfo {title} {{Domain growth morphology in curvature-driven two-dimensional coarsening}},\ }\href {https://doi.org/10.1103/PhysRevE.76.061116} {\bibfield  {journal} {\bibinfo  {journal} {Phys. Rev. E}\ }\textbf {\bibinfo {volume} {76}},\ \bibinfo {pages} {061116} (\bibinfo {year} {2007})}\BibitemShut {NoStop}%
\bibitem [{\citenamefont {Sicilia}\ \emph {et~al.}(2009)\citenamefont {Sicilia}, \citenamefont {Sarrazin}, \citenamefont {Arenzon}, \citenamefont {Bray},\ and\ \citenamefont {Cugliandolo}}]{Sicilia09}%
  \BibitemOpen
  \bibfield  {author} {\bibinfo {author} {\bibfnamefont {A.}~\bibnamefont {Sicilia}}, \bibinfo {author} {\bibfnamefont {Y.}~\bibnamefont {Sarrazin}}, \bibinfo {author} {\bibfnamefont {J.~J.}\ \bibnamefont {Arenzon}}, \bibinfo {author} {\bibfnamefont {A.~J.}\ \bibnamefont {Bray}},\ and\ \bibinfo {author} {\bibfnamefont {L.~F.}\ \bibnamefont {Cugliandolo}},\ }\bibfield  {title} {\bibinfo {title} {Geometry of phase separation},\ }\href {https://doi.org/10.1103/PhysRevE.80.031121} {\bibfield  {journal} {\bibinfo  {journal} {Phys. Rev. E}\ }\textbf {\bibinfo {volume} {80}},\ \bibinfo {pages} {031121} (\bibinfo {year} {2009})}\BibitemShut {NoStop}%
\bibitem [{\citenamefont {Ohring}(2002)}]{Ohring_book02}%
  \BibitemOpen
  \bibfield  {author} {\bibinfo {author} {\bibfnamefont {M.}~\bibnamefont {Ohring}},\ }\href@noop {} {\emph {\bibinfo {title} {Materials science of thin films: depositon and structure}}}\ (\bibinfo  {publisher} {Academic press},\ \bibinfo {year} {2002})\BibitemShut {NoStop}%
\bibitem [{\citenamefont {Barth}\ \emph {et~al.}(2005)\citenamefont {Barth}, \citenamefont {Costantini},\ and\ \citenamefont {Kern}}]{Barth05}%
  \BibitemOpen
  \bibfield  {author} {\bibinfo {author} {\bibfnamefont {J.~V.}\ \bibnamefont {Barth}}, \bibinfo {author} {\bibfnamefont {G.}~\bibnamefont {Costantini}},\ and\ \bibinfo {author} {\bibfnamefont {K.}~\bibnamefont {Kern}},\ }\bibfield  {title} {\bibinfo {title} {{Engineering atomic and molecular nanostructures at surfaces}},\ }\href {https://doi.org/10.1038/nature04166} {\bibfield  {journal} {\bibinfo  {journal} {Nature}\ }\textbf {\bibinfo {volume} {437}},\ \bibinfo {pages} {671} (\bibinfo {year} {2005})}\BibitemShut {NoStop}%
\bibitem [{\citenamefont {Nagel}(2017)}]{Nagel17}%
  \BibitemOpen
  \bibfield  {author} {\bibinfo {author} {\bibfnamefont {S.~R.}\ \bibnamefont {Nagel}},\ }\bibfield  {title} {\bibinfo {title} {Experimental soft-matter science},\ }\href {https://doi.org/10.1103/RevModPhys.89.025002} {\bibfield  {journal} {\bibinfo  {journal} {Rev. Mod. Phys.}\ }\textbf {\bibinfo {volume} {89}},\ \bibinfo {pages} {025002} (\bibinfo {year} {2017})}\BibitemShut {NoStop}%
\bibitem [{\citenamefont {Brú}\ \emph {et~al.}(2003)\citenamefont {Brú}, \citenamefont {Albertos}, \citenamefont {Luis~Subiza}, \citenamefont {García-Asenjo},\ and\ \citenamefont {Brú}}]{Bru03}%
  \BibitemOpen
  \bibfield  {author} {\bibinfo {author} {\bibfnamefont {A.}~\bibnamefont {Brú}}, \bibinfo {author} {\bibfnamefont {S.}~\bibnamefont {Albertos}}, \bibinfo {author} {\bibfnamefont {J.}~\bibnamefont {Luis~Subiza}}, \bibinfo {author} {\bibfnamefont {J.~L.}\ \bibnamefont {García-Asenjo}},\ and\ \bibinfo {author} {\bibfnamefont {I.}~\bibnamefont {Brú}},\ }\bibfield  {title} {\bibinfo {title} {The universal dynamics of tumor growth},\ }\href {https://doi.org/doi: 10.1016/S0006-3495(03)74715-8} {\bibfield  {journal} {\bibinfo  {journal} {Biophys. J.}\ }\textbf {\bibinfo {volume} {85}},\ \bibinfo {pages} {2948} (\bibinfo {year} {2003})}\BibitemShut {NoStop}%
\bibitem [{\citenamefont {Kriecherbauer}\ and\ \citenamefont {Krug}(2010)}]{Kriecherbauer10}%
  \BibitemOpen
  \bibfield  {author} {\bibinfo {author} {\bibfnamefont {T.}~\bibnamefont {Kriecherbauer}}\ and\ \bibinfo {author} {\bibfnamefont {J.}~\bibnamefont {Krug}},\ }\bibfield  {title} {\bibinfo {title} {A pedestrian's view on interacting particle systems, {KPZ} universality and random matrices},\ }\href {https://doi.org/10.1088/1751-8113/43/40/403001} {\bibfield  {journal} {\bibinfo  {journal} {J. Phys. A: Math. Theor.}\ }\textbf {\bibinfo {volume} {43}},\ \bibinfo {pages} {403001} (\bibinfo {year} {2010})}\BibitemShut {NoStop}%
\bibitem [{\citenamefont {Maunuksela}\ \emph {et~al.}(1997)\citenamefont {Maunuksela}, \citenamefont {Myllys}, \citenamefont {K\"ahk\"onen}, \citenamefont {Timonen}, \citenamefont {Provatas}, \citenamefont {Alava},\ and\ \citenamefont {Ala-Nissila}}]{Maunuksela97}%
  \BibitemOpen
  \bibfield  {author} {\bibinfo {author} {\bibfnamefont {J.}~\bibnamefont {Maunuksela}}, \bibinfo {author} {\bibfnamefont {M.}~\bibnamefont {Myllys}}, \bibinfo {author} {\bibfnamefont {O.-P.}\ \bibnamefont {K\"ahk\"onen}}, \bibinfo {author} {\bibfnamefont {J.}~\bibnamefont {Timonen}}, \bibinfo {author} {\bibfnamefont {N.}~\bibnamefont {Provatas}}, \bibinfo {author} {\bibfnamefont {M.~J.}\ \bibnamefont {Alava}},\ and\ \bibinfo {author} {\bibfnamefont {T.}~\bibnamefont {Ala-Nissila}},\ }\bibfield  {title} {\bibinfo {title} {Kinetic roughening in slow combustion of paper},\ }\href {https://doi.org/10.1103/PhysRevLett.79.1515} {\bibfield  {journal} {\bibinfo  {journal} {Phys. Rev. Lett.}\ }\textbf {\bibinfo {volume} {79}},\ \bibinfo {pages} {1515} (\bibinfo {year} {1997})}\BibitemShut {NoStop}%
\bibitem [{\citenamefont {Takeuchi}\ and\ \citenamefont {Sano}(2010)}]{Takeuchi10}%
  \BibitemOpen
  \bibfield  {author} {\bibinfo {author} {\bibfnamefont {K.~A.}\ \bibnamefont {Takeuchi}}\ and\ \bibinfo {author} {\bibfnamefont {M.}~\bibnamefont {Sano}},\ }\bibfield  {title} {\bibinfo {title} {Universal fluctuations of growing interfaces: Evidence in turbulent liquid crystals},\ }\href {https://doi.org/10.1103/PhysRevLett.104.230601} {\bibfield  {journal} {\bibinfo  {journal} {Phys. Rev. Lett.}\ }\textbf {\bibinfo {volume} {104}},\ \bibinfo {pages} {230601} (\bibinfo {year} {2010})}\BibitemShut {NoStop}%
\bibitem [{\citenamefont {Ojeda}\ \emph {et~al.}(2000)\citenamefont {Ojeda}, \citenamefont {Cuerno}, \citenamefont {Salvarezza},\ and\ \citenamefont {V\'azquez}}]{Ojeda00}%
  \BibitemOpen
  \bibfield  {author} {\bibinfo {author} {\bibfnamefont {F.}~\bibnamefont {Ojeda}}, \bibinfo {author} {\bibfnamefont {R.}~\bibnamefont {Cuerno}}, \bibinfo {author} {\bibfnamefont {R.}~\bibnamefont {Salvarezza}},\ and\ \bibinfo {author} {\bibfnamefont {L.}~\bibnamefont {V\'azquez}},\ }\bibfield  {title} {\bibinfo {title} {Dynamics of rough interfaces in chemical vapor deposition: Experiments and a model for silica films},\ }\href {https://doi.org/10.1103/PhysRevLett.84.3125} {\bibfield  {journal} {\bibinfo  {journal} {Phys. Rev. Lett.}\ }\textbf {\bibinfo {volume} {84}},\ \bibinfo {pages} {3125} (\bibinfo {year} {2000})}\BibitemShut {NoStop}%
\bibitem [{\citenamefont {Palasantzas}\ \emph {et~al.}(2002)\citenamefont {Palasantzas}, \citenamefont {Tsamouras},\ and\ \citenamefont {{De Hosson}}}]{Palasantzas02}%
  \BibitemOpen
  \bibfield  {author} {\bibinfo {author} {\bibfnamefont {G.}~\bibnamefont {Palasantzas}}, \bibinfo {author} {\bibfnamefont {D.}~\bibnamefont {Tsamouras}},\ and\ \bibinfo {author} {\bibfnamefont {J.}~\bibnamefont {{De Hosson}}},\ }\bibfield  {title} {\bibinfo {title} {Roughening aspects of room temperature vapor deposited oligomer thin films onto si substrates},\ }\href {https://doi.org/https://doi.org/10.1016/S0039-6028(02)01271-2} {\bibfield  {journal} {\bibinfo  {journal} {Surface Science}\ }\textbf {\bibinfo {volume} {507-510}},\ \bibinfo {pages} {357} (\bibinfo {year} {2002})}\BibitemShut {NoStop}%
\bibitem [{\citenamefont {Almeida}\ \emph {et~al.}(2014)\citenamefont {Almeida}, \citenamefont {Ferreira}, \citenamefont {Oliveira},\ and\ \citenamefont {Aar{\~a}o~Reis}}]{Almeida14}%
  \BibitemOpen
  \bibfield  {author} {\bibinfo {author} {\bibfnamefont {R.~A.~L.}\ \bibnamefont {Almeida}}, \bibinfo {author} {\bibfnamefont {S.~O.}\ \bibnamefont {Ferreira}}, \bibinfo {author} {\bibfnamefont {T.~J.}\ \bibnamefont {Oliveira}},\ and\ \bibinfo {author} {\bibfnamefont {F.~D.~A.}\ \bibnamefont {Aar{\~a}o~Reis}},\ }\bibfield  {title} {\bibinfo {title} {Universal fluctuations in the growth of semiconductor thin films},\ }\href {https://doi.org/10.1103/PhysRevB.89.045309} {\bibfield  {journal} {\bibinfo  {journal} {Phys. Rev. B}\ }\textbf {\bibinfo {volume} {89}},\ \bibinfo {pages} {045309} (\bibinfo {year} {2014})}\BibitemShut {NoStop}%
\bibitem [{\citenamefont {Halpin-Healy}\ and\ \citenamefont {Palasantzas}(2014)}]{HHealy14}%
  \BibitemOpen
  \bibfield  {author} {\bibinfo {author} {\bibfnamefont {T.}~\bibnamefont {Halpin-Healy}}\ and\ \bibinfo {author} {\bibfnamefont {G.}~\bibnamefont {Palasantzas}},\ }\bibfield  {title} {\bibinfo {title} {Universal correlators and distributions as experimental signatures of (2 + 1)-dimensional {K}ardar-{P}arisi-{Z}hang growth},\ }\href {https://doi.org/10.1209/0295-5075/105/50001} {\bibfield  {journal} {\bibinfo  {journal} {Europhys. Lett.}\ }\textbf {\bibinfo {volume} {105}},\ \bibinfo {pages} {50001} (\bibinfo {year} {2014})}\BibitemShut {NoStop}%
\bibitem [{\citenamefont {Kondev}\ and\ \citenamefont {Henley}(1995)}]{Kondev95}%
  \BibitemOpen
  \bibfield  {author} {\bibinfo {author} {\bibfnamefont {J.}~\bibnamefont {Kondev}}\ and\ \bibinfo {author} {\bibfnamefont {C.~L.}\ \bibnamefont {Henley}},\ }\bibfield  {title} {\bibinfo {title} {Geometrical exponents of contour loops on random {G}aussian surfaces},\ }\href {https://doi.org/10.1103/PhysRevLett.74.4580} {\bibfield  {journal} {\bibinfo  {journal} {Phys. Rev. Lett.}\ }\textbf {\bibinfo {volume} {74}},\ \bibinfo {pages} {4580} (\bibinfo {year} {1995})}\BibitemShut {NoStop}%
\bibitem [{\citenamefont {Olami}\ and\ \citenamefont {Zeitak}(1996)}]{Olami96}%
  \BibitemOpen
  \bibfield  {author} {\bibinfo {author} {\bibfnamefont {Z.}~\bibnamefont {Olami}}\ and\ \bibinfo {author} {\bibfnamefont {R.}~\bibnamefont {Zeitak}},\ }\bibfield  {title} {\bibinfo {title} {Scaling of island distributions, percolation, and criticality in contour cuts through wrinkled surfaces},\ }\href {https://doi.org/10.1103/PhysRevLett.76.247} {\bibfield  {journal} {\bibinfo  {journal} {Phys. Rev. Lett.}\ }\textbf {\bibinfo {volume} {76}},\ \bibinfo {pages} {247} (\bibinfo {year} {1996})}\BibitemShut {NoStop}%
\bibitem [{\citenamefont {Schwartz}(2001)}]{Schwartz01}%
  \BibitemOpen
  \bibfield  {author} {\bibinfo {author} {\bibfnamefont {M.}~\bibnamefont {Schwartz}},\ }\bibfield  {title} {\bibinfo {title} {End-to-end distance on contour loops of random {G}aussian surfaces},\ }\href {https://doi.org/10.1103/PhysRevLett.86.1283} {\bibfield  {journal} {\bibinfo  {journal} {Phys. Rev. Lett.}\ }\textbf {\bibinfo {volume} {86}},\ \bibinfo {pages} {1283} (\bibinfo {year} {2001})}\BibitemShut {NoStop}%
\bibitem [{\citenamefont {Schramm}\ and\ \citenamefont {Sheffield}(2009)}]{Schramm09}%
  \BibitemOpen
  \bibfield  {author} {\bibinfo {author} {\bibfnamefont {O.}~\bibnamefont {Schramm}}\ and\ \bibinfo {author} {\bibfnamefont {S.}~\bibnamefont {Sheffield}},\ }\bibfield  {title} {\bibinfo {title} {{Contour lines of the two-dimensional discrete {G}aussian free field}},\ }\href {https://doi.org/10.1007/s11511-009-0034-y} {\bibfield  {journal} {\bibinfo  {journal} {Acta Math.}\ }\textbf {\bibinfo {volume} {202}},\ \bibinfo {pages} {21 } (\bibinfo {year} {2009})}\BibitemShut {NoStop}%
\bibitem [{\citenamefont {Giordanelli}\ \emph {et~al.}(2016)\citenamefont {Giordanelli}, \citenamefont {Posé}, \citenamefont {Mendoza},\ and\ \citenamefont {Herrmann}}]{Giordanelli16}%
  \BibitemOpen
  \bibfield  {author} {\bibinfo {author} {\bibfnamefont {I.}~\bibnamefont {Giordanelli}}, \bibinfo {author} {\bibfnamefont {N.}~\bibnamefont {Posé}}, \bibinfo {author} {\bibfnamefont {M.}~\bibnamefont {Mendoza}},\ and\ \bibinfo {author} {\bibfnamefont {H.~J.}\ \bibnamefont {Herrmann}},\ }\bibfield  {title} {\bibinfo {title} {Conformal invariance of graphene sheets},\ }\href {https://doi.org/10.1038/srep22949} {\bibfield  {journal} {\bibinfo  {journal} {Sci. Rep.}\ }\textbf {\bibinfo {volume} {6}},\ \bibinfo {pages} {22949} (\bibinfo {year} {2016})}\BibitemShut {NoStop}%
\bibitem [{\citenamefont {Saberi}\ \emph {et~al.}(2008)\citenamefont {Saberi}, \citenamefont {Niry}, \citenamefont {Fazeli}, \citenamefont {Rahimi~Tabar},\ and\ \citenamefont {Rouhani}}]{Saberi08}%
  \BibitemOpen
  \bibfield  {author} {\bibinfo {author} {\bibfnamefont {A.~A.}\ \bibnamefont {Saberi}}, \bibinfo {author} {\bibfnamefont {M.~D.}\ \bibnamefont {Niry}}, \bibinfo {author} {\bibfnamefont {S.~M.}\ \bibnamefont {Fazeli}}, \bibinfo {author} {\bibfnamefont {M.~R.}\ \bibnamefont {Rahimi~Tabar}},\ and\ \bibinfo {author} {\bibfnamefont {S.}~\bibnamefont {Rouhani}},\ }\bibfield  {title} {\bibinfo {title} {Conformal invariance of isoheight lines in a two-dimensional {K}ardar-{P}arisi-{Z}hang surface},\ }\href {https://doi.org/10.1103/PhysRevE.77.051607} {\bibfield  {journal} {\bibinfo  {journal} {Phys. Rev. E}\ }\textbf {\bibinfo {volume} {77}},\ \bibinfo {pages} {051607} (\bibinfo {year} {2008})}\BibitemShut {NoStop}%
\bibitem [{\citenamefont {Saberi}\ and\ \citenamefont {Rouhani}(2009)}]{Saberi09}%
  \BibitemOpen
  \bibfield  {author} {\bibinfo {author} {\bibfnamefont {A.~A.}\ \bibnamefont {Saberi}}\ and\ \bibinfo {author} {\bibfnamefont {S.}~\bibnamefont {Rouhani}},\ }\bibfield  {title} {\bibinfo {title} {Scaling of clusters and winding-angle statistics of isoheight lines in two-dimensional {K}ardar-{P}arisi-{Z}hang surfaces},\ }\href {https://doi.org/10.1103/PhysRevE.79.036102} {\bibfield  {journal} {\bibinfo  {journal} {Phys. Rev. E}\ }\textbf {\bibinfo {volume} {79}},\ \bibinfo {pages} {036102} (\bibinfo {year} {2009})}\BibitemShut {NoStop}%
\bibitem [{\citenamefont {Saberi}\ \emph {et~al.}(2010)\citenamefont {Saberi}, \citenamefont {Dashti-Naserabadi},\ and\ \citenamefont {Rouhani}}]{Saberi10}%
  \BibitemOpen
  \bibfield  {author} {\bibinfo {author} {\bibfnamefont {A.~A.}\ \bibnamefont {Saberi}}, \bibinfo {author} {\bibfnamefont {H.}~\bibnamefont {Dashti-Naserabadi}},\ and\ \bibinfo {author} {\bibfnamefont {S.}~\bibnamefont {Rouhani}},\ }\bibfield  {title} {\bibinfo {title} {Classification of $(2+1)$-dimensional growing surfaces using {S}chramm-{L}oewner evolution},\ }\href {https://doi.org/10.1103/PhysRevE.82.020101} {\bibfield  {journal} {\bibinfo  {journal} {Phys. Rev. E}\ }\textbf {\bibinfo {volume} {82}},\ \bibinfo {pages} {020101} (\bibinfo {year} {2010})}\BibitemShut {NoStop}%
\bibitem [{\citenamefont {Kardar}\ \emph {et~al.}(1986)\citenamefont {Kardar}, \citenamefont {Parisi},\ and\ \citenamefont {Zhang}}]{Kardar86}%
  \BibitemOpen
  \bibfield  {author} {\bibinfo {author} {\bibfnamefont {M.}~\bibnamefont {Kardar}}, \bibinfo {author} {\bibfnamefont {G.}~\bibnamefont {Parisi}},\ and\ \bibinfo {author} {\bibfnamefont {Y.-C.}\ \bibnamefont {Zhang}},\ }\bibfield  {title} {\bibinfo {title} {{Dynamic Scaling of Growing Interfaces}},\ }\href {https://doi.org/10.1103/PhysRevLett.56.889} {\bibfield  {journal} {\bibinfo  {journal} {Phys. Rev. Lett.}\ }\textbf {\bibinfo {volume} {56}},\ \bibinfo {pages} {889} (\bibinfo {year} {1986})}\BibitemShut {NoStop}%
\bibitem [{\citenamefont {Edwards}\ and\ \citenamefont {Wilkinson}(1982)}]{Edwards82}%
  \BibitemOpen
  \bibfield  {author} {\bibinfo {author} {\bibfnamefont {S.~F.}\ \bibnamefont {Edwards}}\ and\ \bibinfo {author} {\bibfnamefont {D.~R.}\ \bibnamefont {Wilkinson}},\ }\bibfield  {title} {\bibinfo {title} {The surface statistics of a granular aggregate},\ }\href {https://doi.org/10.1098/rspa.1982.0056} {\bibfield  {journal} {\bibinfo  {journal} {Proc. of the Royal Soc. of London. A. Math. and Phys. Sci.}\ }\textbf {\bibinfo {volume} {381}},\ \bibinfo {pages} {17} (\bibinfo {year} {1982})}\BibitemShut {NoStop}%
\bibitem [{\citenamefont {Krug}\ and\ \citenamefont {Spohn}(1991)}]{Krug91}%
  \BibitemOpen
  \bibfield  {author} {\bibinfo {author} {\bibfnamefont {J.}~\bibnamefont {Krug}}\ and\ \bibinfo {author} {\bibfnamefont {H.}~\bibnamefont {Spohn}},\ }\bibfield  {title} {\bibinfo {title} {Kinetic roughening of growing surfaces},\ }in\ \href@noop {} {\emph {\bibinfo {booktitle} {Solids far from equilibrium}}},\ \bibinfo {editor} {edited by\ \bibinfo {editor} {\bibfnamefont {C.}~\bibnamefont {Godr\`eche}}}\ (\bibinfo  {publisher} {Cambridge University Press Cambridge},\ \bibinfo {year} {1991})\BibitemShut {NoStop}%
\bibitem [{\citenamefont {Corwin}(2012)}]{Corwin11}%
  \BibitemOpen
  \bibfield  {author} {\bibinfo {author} {\bibfnamefont {I.}~\bibnamefont {Corwin}},\ }\bibfield  {title} {\bibinfo {title} {The {K}ardar-{P}arisi-{Z}hang equation and universality class},\ }\href {https://doi.org/10.1142/S2010326311300014} {\bibfield  {journal} {\bibinfo  {journal} {Random Matrices: Theory Appl.}\ }\textbf {\bibinfo {volume} {01}},\ \bibinfo {pages} {1130001} (\bibinfo {year} {2012})}\BibitemShut {NoStop}%
\bibitem [{\citenamefont {Halpin-Healy}\ and\ \citenamefont {Takeuchi}(2015)}]{HHealy15}%
  \BibitemOpen
  \bibfield  {author} {\bibinfo {author} {\bibfnamefont {T.}~\bibnamefont {Halpin-Healy}}\ and\ \bibinfo {author} {\bibfnamefont {K.~A.}\ \bibnamefont {Takeuchi}},\ }\bibfield  {title} {\bibinfo {title} {{A KPZ Cocktail-Shaken, not Stirred...}},\ }\href {https://doi.org/10.1007/s10955-015-1282-1} {\bibfield  {journal} {\bibinfo  {journal} {J. Stat. Phys.}\ }\textbf {\bibinfo {volume} {160}},\ \bibinfo {pages} {794} (\bibinfo {year} {2015})}\BibitemShut {NoStop}%
\bibitem [{\citenamefont {Takeuchi}(2018)}]{Takeuchi18}%
  \BibitemOpen
  \bibfield  {author} {\bibinfo {author} {\bibfnamefont {K.~A.}\ \bibnamefont {Takeuchi}},\ }\bibfield  {title} {\bibinfo {title} {An appetizer to modern developments on the {K}ardar–{P}arisi–{Z}hang universality class},\ }\href {https://doi.org/https://doi.org/10.1016/j.physa.2018.03.009} {\bibfield  {journal} {\bibinfo  {journal} {Physica A}\ }\textbf {\bibinfo {volume} {504}},\ \bibinfo {pages} {77} (\bibinfo {year} {2018})}\BibitemShut {NoStop}%
\bibitem [{\citenamefont {Nahum}\ \emph {et~al.}(2017)\citenamefont {Nahum}, \citenamefont {Ruhman}, \citenamefont {Vijay},\ and\ \citenamefont {Haah}}]{Nahum17}%
  \BibitemOpen
  \bibfield  {author} {\bibinfo {author} {\bibfnamefont {A.}~\bibnamefont {Nahum}}, \bibinfo {author} {\bibfnamefont {J.}~\bibnamefont {Ruhman}}, \bibinfo {author} {\bibfnamefont {S.}~\bibnamefont {Vijay}},\ and\ \bibinfo {author} {\bibfnamefont {J.}~\bibnamefont {Haah}},\ }\bibfield  {title} {\bibinfo {title} {Quantum entanglement growth under random unitary dynamics},\ }\href {https://doi.org/10.1103/PhysRevX.7.031016} {\bibfield  {journal} {\bibinfo  {journal} {Phys. Rev. X}\ }\textbf {\bibinfo {volume} {7}},\ \bibinfo {pages} {031016} (\bibinfo {year} {2017})}\BibitemShut {NoStop}%
\bibitem [{\citenamefont {Fontaine}\ \emph {et~al.}(2022)\citenamefont {Fontaine}, \citenamefont {Squizzato}, \citenamefont {Baboux}, \citenamefont {Amelio}, \citenamefont {Lemaître}, \citenamefont {Morassi}, \citenamefont {Sagnes}, \citenamefont {Le~Gratiet}, \citenamefont {Harouri}, \citenamefont {Wouters}, \citenamefont {Carusotto}, \citenamefont {Amo}, \citenamefont {Richard}, \citenamefont {Minguzzi}, \citenamefont {Canet}, \citenamefont {Ravets},\ and\ \citenamefont {Bloch}}]{Fontaine22}%
  \BibitemOpen
  \bibfield  {author} {\bibinfo {author} {\bibfnamefont {Q.}~\bibnamefont {Fontaine}}, \bibinfo {author} {\bibfnamefont {D.}~\bibnamefont {Squizzato}}, \bibinfo {author} {\bibfnamefont {F.}~\bibnamefont {Baboux}}, \bibinfo {author} {\bibfnamefont {I.}~\bibnamefont {Amelio}}, \bibinfo {author} {\bibfnamefont {A.}~\bibnamefont {Lemaître}}, \bibinfo {author} {\bibfnamefont {M.}~\bibnamefont {Morassi}}, \bibinfo {author} {\bibfnamefont {I.}~\bibnamefont {Sagnes}}, \bibinfo {author} {\bibfnamefont {L.}~\bibnamefont {Le~Gratiet}}, \bibinfo {author} {\bibfnamefont {A.}~\bibnamefont {Harouri}}, \bibinfo {author} {\bibfnamefont {M.}~\bibnamefont {Wouters}}, \bibinfo {author} {\bibfnamefont {I.}~\bibnamefont {Carusotto}}, \bibinfo {author} {\bibfnamefont {A.}~\bibnamefont {Amo}}, \bibinfo {author} {\bibfnamefont {M.}~\bibnamefont {Richard}}, \bibinfo {author} {\bibfnamefont {A.}~\bibnamefont {Minguzzi}}, \bibinfo {author} {\bibfnamefont {L.}~\bibnamefont {Canet}}, \bibinfo {author} {\bibfnamefont {S.}~\bibnamefont
  {Ravets}},\ and\ \bibinfo {author} {\bibfnamefont {J.}~\bibnamefont {Bloch}},\ }\bibfield  {title} {\bibinfo {title} {{K}ardar–{P}arisi–{Z}hang universality in a one-dimensional polariton condensate},\ }\href {https://doi.org/10.1038/s41586-022-05001-8} {\bibfield  {journal} {\bibinfo  {journal} {Nature}\ }\textbf {\bibinfo {volume} {608}},\ \bibinfo {pages} {687} (\bibinfo {year} {2022})}\BibitemShut {NoStop}%
\bibitem [{\citenamefont {Widmann}\ \emph {et~al.}(2025)\citenamefont {Widmann}, \citenamefont {Dam}, \citenamefont {Düreth}, \citenamefont {Mayer}, \citenamefont {Daviet}, \citenamefont {Zelle}, \citenamefont {Laibacher}, \citenamefont {Emmerling}, \citenamefont {Kamp}, \citenamefont {Diehl}, \citenamefont {Betzold}, \citenamefont {Klembt},\ and\ \citenamefont {Höfling}}]{Widmann25}%
  \BibitemOpen
  \bibfield  {author} {\bibinfo {author} {\bibfnamefont {S.}~\bibnamefont {Widmann}}, \bibinfo {author} {\bibfnamefont {S.}~\bibnamefont {Dam}}, \bibinfo {author} {\bibfnamefont {J.}~\bibnamefont {Düreth}}, \bibinfo {author} {\bibfnamefont {C.~G.}\ \bibnamefont {Mayer}}, \bibinfo {author} {\bibfnamefont {R.}~\bibnamefont {Daviet}}, \bibinfo {author} {\bibfnamefont {C.~P.}\ \bibnamefont {Zelle}}, \bibinfo {author} {\bibfnamefont {D.}~\bibnamefont {Laibacher}}, \bibinfo {author} {\bibfnamefont {M.}~\bibnamefont {Emmerling}}, \bibinfo {author} {\bibfnamefont {M.}~\bibnamefont {Kamp}}, \bibinfo {author} {\bibfnamefont {S.}~\bibnamefont {Diehl}}, \bibinfo {author} {\bibfnamefont {S.}~\bibnamefont {Betzold}}, \bibinfo {author} {\bibfnamefont {S.}~\bibnamefont {Klembt}},\ and\ \bibinfo {author} {\bibfnamefont {S.}~\bibnamefont {Höfling}},\ }\href {https://arxiv.org/abs/2506.15521} {\bibinfo {title} {Observation of {K}ardar-{P}arisi-{Z}hang universal scaling in two dimensions}} (\bibinfo {year} {2025}),\ \Eprint
  {https://arxiv.org/abs/2506.15521} {arXiv:2506.15521 [quant-ph]} \BibitemShut {NoStop}%
\bibitem [{\citenamefont {Ljubotina}\ \emph {et~al.}(2019)\citenamefont {Ljubotina}, \citenamefont {\ifmmode \check{Z}\else \v{Z}\fi{}nidari\ifmmode~\check{c}\else \v{c}\fi{}},\ and\ \citenamefont {Prosen}}]{Ljubotina19}%
  \BibitemOpen
  \bibfield  {author} {\bibinfo {author} {\bibfnamefont {M.}~\bibnamefont {Ljubotina}}, \bibinfo {author} {\bibfnamefont {M.}~\bibnamefont {\ifmmode \check{Z}\else \v{Z}\fi{}nidari\ifmmode~\check{c}\else \v{c}\fi{}}},\ and\ \bibinfo {author} {\bibfnamefont {T.~c.~v.}\ \bibnamefont {Prosen}},\ }\bibfield  {title} {\bibinfo {title} {{K}ardar-{P}arisi-{Z}hang physics in the quantum {H}eisenberg magnet},\ }\href {https://doi.org/10.1103/PhysRevLett.122.210602} {\bibfield  {journal} {\bibinfo  {journal} {Phys. Rev. Lett.}\ }\textbf {\bibinfo {volume} {122}},\ \bibinfo {pages} {210602} (\bibinfo {year} {2019})}\BibitemShut {NoStop}%
\bibitem [{\citenamefont {Ye}\ \emph {et~al.}(2022)\citenamefont {Ye}, \citenamefont {Machado}, \citenamefont {Kemp}, \citenamefont {Hutson},\ and\ \citenamefont {Yao}}]{Ye22}%
  \BibitemOpen
  \bibfield  {author} {\bibinfo {author} {\bibfnamefont {B.}~\bibnamefont {Ye}}, \bibinfo {author} {\bibfnamefont {F.}~\bibnamefont {Machado}}, \bibinfo {author} {\bibfnamefont {J.}~\bibnamefont {Kemp}}, \bibinfo {author} {\bibfnamefont {R.~B.}\ \bibnamefont {Hutson}},\ and\ \bibinfo {author} {\bibfnamefont {N.~Y.}\ \bibnamefont {Yao}},\ }\bibfield  {title} {\bibinfo {title} {Universal {K}ardar–{P}arisi–{Z}hang dynamics in integrable quantum systems},\ }\href {https://doi.org/10.1103/PhysRevLett.129.230602} {\bibfield  {journal} {\bibinfo  {journal} {Phys. Rev. Lett.}\ }\textbf {\bibinfo {volume} {129}},\ \bibinfo {pages} {230602} (\bibinfo {year} {2022})}\BibitemShut {NoStop}%
\bibitem [{\citenamefont {Takeuchi}\ \emph {et~al.}(2025)\citenamefont {Takeuchi}, \citenamefont {Takasan}, \citenamefont {Busani}, \citenamefont {Ferrari}, \citenamefont {Vasseur},\ and\ \citenamefont {De~Nardis}}]{Takeuchi25}%
  \BibitemOpen
  \bibfield  {author} {\bibinfo {author} {\bibfnamefont {K.~A.}\ \bibnamefont {Takeuchi}}, \bibinfo {author} {\bibfnamefont {K.}~\bibnamefont {Takasan}}, \bibinfo {author} {\bibfnamefont {O.}~\bibnamefont {Busani}}, \bibinfo {author} {\bibfnamefont {P.~L.}\ \bibnamefont {Ferrari}}, \bibinfo {author} {\bibfnamefont {R.}~\bibnamefont {Vasseur}},\ and\ \bibinfo {author} {\bibfnamefont {J.}~\bibnamefont {De~Nardis}},\ }\bibfield  {title} {\bibinfo {title} {Partial yet definite emergence of the {K}ardar–{P}arisi–{Z}hang class in isotropic spin chains},\ }\href {https://doi.org/10.1103/PhysRevLett.134.097104} {\bibfield  {journal} {\bibinfo  {journal} {Phys. Rev. Lett.}\ }\textbf {\bibinfo {volume} {134}},\ \bibinfo {pages} {097104} (\bibinfo {year} {2025})}\BibitemShut {NoStop}%
\bibitem [{\citenamefont {Nattermann}\ and\ \citenamefont {Tang}(1992)}]{Nattermann92}%
  \BibitemOpen
  \bibfield  {author} {\bibinfo {author} {\bibfnamefont {T.}~\bibnamefont {Nattermann}}\ and\ \bibinfo {author} {\bibfnamefont {L.-H.}\ \bibnamefont {Tang}},\ }\bibfield  {title} {\bibinfo {title} {Kinetic surface roughening. i. the {K}ardar-{P}arisi-{Z}hang equation in the weak-coupling regime},\ }\href {https://doi.org/10.1103/PhysRevA.45.7156} {\bibfield  {journal} {\bibinfo  {journal} {Phys. Rev. A}\ }\textbf {\bibinfo {volume} {45}},\ \bibinfo {pages} {7156} (\bibinfo {year} {1992})}\BibitemShut {NoStop}%
\bibitem [{Note1()}]{Note1}%
  \BibitemOpen
  \bibinfo {note} {Some values predicted by mode-coupling~\cite {Doherty94, *Tu94, *Colaiori01}, operator product expansion~\cite {Lassig98}, nonperturbative renormalization group~\cite {Canet10, *Kloss12}, and conjectures~\cite {Wolf87, Kim89, Oliveira22} are fairly close to recent numerical evaluations~\cite {Kelling11, *Kelling18, Pagnani15}.}\BibitemShut {Stop}%
\bibitem [{\citenamefont {Halpin-Healy}(2012)}]{Hhealy12}%
  \BibitemOpen
  \bibfield  {author} {\bibinfo {author} {\bibfnamefont {T.}~\bibnamefont {Halpin-Healy}},\ }\bibfield  {title} {\bibinfo {title} {($2\mathbf{+}1$)-dimensional directed polymer in a random medium: Scaling phenomena and universal distributions},\ }\href {https://doi.org/10.1103/PhysRevLett.109.170602} {\bibfield  {journal} {\bibinfo  {journal} {Phys. Rev. Lett.}\ }\textbf {\bibinfo {volume} {109}},\ \bibinfo {pages} {170602} (\bibinfo {year} {2012})}\BibitemShut {NoStop}%
\bibitem [{\citenamefont {Halpin-Healy}(2013)}]{HHealy13}%
  \BibitemOpen
  \bibfield  {author} {\bibinfo {author} {\bibfnamefont {T.}~\bibnamefont {Halpin-Healy}},\ }\bibfield  {title} {\bibinfo {title} {Extremal paths, the stochastic heat equation, and the three-dimensional {K}ardar-{P}arisi-{Z}hang universality class},\ }\href {https://doi.org/10.1103/PhysRevE.88.042118} {\bibfield  {journal} {\bibinfo  {journal} {Phys. Rev. E}\ }\textbf {\bibinfo {volume} {88}},\ \bibinfo {pages} {042118} (\bibinfo {year} {2013})}\BibitemShut {NoStop}%
\bibitem [{\citenamefont {Carrasco}\ \emph {et~al.}(2014)\citenamefont {Carrasco}, \citenamefont {Takeuchi}, \citenamefont {Ferreira},\ and\ \citenamefont {Oliveira}}]{Carrasco14}%
  \BibitemOpen
  \bibfield  {author} {\bibinfo {author} {\bibfnamefont {I.~S.~S.}\ \bibnamefont {Carrasco}}, \bibinfo {author} {\bibfnamefont {K.~A.}\ \bibnamefont {Takeuchi}}, \bibinfo {author} {\bibfnamefont {S.~C.}\ \bibnamefont {Ferreira}},\ and\ \bibinfo {author} {\bibfnamefont {T.~J.}\ \bibnamefont {Oliveira}},\ }\bibfield  {title} {\bibinfo {title} {Interface fluctuations for deposition on enlarging flat substrates},\ }\href {https://doi.org/10.1088/1367-2630/16/12/123057} {\bibfield  {journal} {\bibinfo  {journal} {New J. Phys.}\ }\textbf {\bibinfo {volume} {16}},\ \bibinfo {pages} {123057} (\bibinfo {year} {2014})}\BibitemShut {NoStop}%
\bibitem [{\citenamefont {Carrasco}\ and\ \citenamefont {Oliveira}(2022)}]{Carrasco22}%
  \BibitemOpen
  \bibfield  {author} {\bibinfo {author} {\bibfnamefont {I.~S.~S.}\ \bibnamefont {Carrasco}}\ and\ \bibinfo {author} {\bibfnamefont {T.~J.}\ \bibnamefont {Oliveira}},\ }\bibfield  {title} {\bibinfo {title} {{K}ardar-{P}arisi-{Z}hang growth on square domains that enlarge nonlinearly in time},\ }\href {https://doi.org/10.1103/PhysRevE.105.054804} {\bibfield  {journal} {\bibinfo  {journal} {Phys. Rev. E}\ }\textbf {\bibinfo {volume} {105}},\ \bibinfo {pages} {054804} (\bibinfo {year} {2022})}\BibitemShut {NoStop}%
\bibitem [{\citenamefont {Carrasco}\ and\ \citenamefont {Oliveira}(2023)}]{Carrasco23}%
  \BibitemOpen
  \bibfield  {author} {\bibinfo {author} {\bibfnamefont {I.~S.~S.}\ \bibnamefont {Carrasco}}\ and\ \bibinfo {author} {\bibfnamefont {T.~J.}\ \bibnamefont {Oliveira}},\ }\bibfield  {title} {\bibinfo {title} {One-point height fluctuations and two-point correlators of $(2+1)$ cylindrical {K}{P}{Z} systems},\ }\href {https://doi.org/10.1103/PhysRevE.107.064140} {\bibfield  {journal} {\bibinfo  {journal} {Phys. Rev. E}\ }\textbf {\bibinfo {volume} {107}},\ \bibinfo {pages} {064140} (\bibinfo {year} {2023})}\BibitemShut {NoStop}%
\bibitem [{\citenamefont {Newman}\ and\ \citenamefont {Swift}(1997)}]{Newman97}%
  \BibitemOpen
  \bibfield  {author} {\bibinfo {author} {\bibfnamefont {T.~J.}\ \bibnamefont {Newman}}\ and\ \bibinfo {author} {\bibfnamefont {M.~R.}\ \bibnamefont {Swift}},\ }\bibfield  {title} {\bibinfo {title} {Nonuniversal exponents in interface growth},\ }\href {https://doi.org/10.1103/PhysRevLett.79.2261} {\bibfield  {journal} {\bibinfo  {journal} {Phys. Rev. Lett.}\ }\textbf {\bibinfo {volume} {79}},\ \bibinfo {pages} {2261} (\bibinfo {year} {1997})}\BibitemShut {NoStop}%
\bibitem [{\citenamefont {Moser}\ \emph {et~al.}(1991)\citenamefont {Moser}, \citenamefont {Kertész},\ and\ \citenamefont {Wolf}}]{Moser91}%
  \BibitemOpen
  \bibfield  {author} {\bibinfo {author} {\bibfnamefont {K.}~\bibnamefont {Moser}}, \bibinfo {author} {\bibfnamefont {J.}~\bibnamefont {Kertész}},\ and\ \bibinfo {author} {\bibfnamefont {D.~E.}\ \bibnamefont {Wolf}},\ }\bibfield  {title} {\bibinfo {title} {Numerical solution of the {K}ardar-{P}arisi-{Z}hang equation in one, two and three dimensions},\ }\href {https://doi.org/https://doi.org/10.1016/0378-4371(91)90017-7} {\bibfield  {journal} {\bibinfo  {journal} {Physica A}\ }\textbf {\bibinfo {volume} {178}},\ \bibinfo {pages} {215} (\bibinfo {year} {1991})}\BibitemShut {NoStop}%
\bibitem [{\citenamefont {Kim}\ and\ \citenamefont {Kosterlitz}(1989)}]{Kim89}%
  \BibitemOpen
  \bibfield  {author} {\bibinfo {author} {\bibfnamefont {J.~M.}\ \bibnamefont {Kim}}\ and\ \bibinfo {author} {\bibfnamefont {J.~M.}\ \bibnamefont {Kosterlitz}},\ }\bibfield  {title} {\bibinfo {title} {Growth in a restricted solid-on-solid model},\ }\href {https://doi.org/10.1103/PhysRevLett.62.2289} {\bibfield  {journal} {\bibinfo  {journal} {Phys. Rev. Lett.}\ }\textbf {\bibinfo {volume} {62}},\ \bibinfo {pages} {2289} (\bibinfo {year} {1989})}\BibitemShut {NoStop}%
\bibitem [{\citenamefont {Mello}\ \emph {et~al.}(2001)\citenamefont {Mello}, \citenamefont {Chaves},\ and\ \citenamefont {Oliveira}}]{Mello01}%
  \BibitemOpen
  \bibfield  {author} {\bibinfo {author} {\bibfnamefont {B.~A.}\ \bibnamefont {Mello}}, \bibinfo {author} {\bibfnamefont {A.~S.}\ \bibnamefont {Chaves}},\ and\ \bibinfo {author} {\bibfnamefont {F.~A.}\ \bibnamefont {Oliveira}},\ }\bibfield  {title} {\bibinfo {title} {Discrete atomistic model to simulate etching of a crystalline solid},\ }\href {https://doi.org/10.1103/PhysRevE.63.041113} {\bibfield  {journal} {\bibinfo  {journal} {Phys. Rev. E}\ }\textbf {\bibinfo {volume} {63}},\ \bibinfo {pages} {041113} (\bibinfo {year} {2001})}\BibitemShut {NoStop}%
\bibitem [{\citenamefont {Kim}(2015)}]{Kim15}%
  \BibitemOpen
  \bibfield  {author} {\bibinfo {author} {\bibfnamefont {J.~M.}\ \bibnamefont {Kim}},\ }\bibfield  {title} {\bibinfo {title} {{Restricted solid-on-solid model in d = 2 + 1 dimension with various restriction parameters N}},\ }\href {https://doi.org/10.3938/jkps.67.1529} {\bibfield  {journal} {\bibinfo  {journal} {J. Korean Phys. Soc.}\ }\textbf {\bibinfo {volume} {67}},\ \bibinfo {pages} {1529} (\bibinfo {year} {2015})}\BibitemShut {NoStop}%
\bibitem [{\citenamefont {Family}\ and\ \citenamefont {Vicsek}(1985)}]{Family85}%
  \BibitemOpen
  \bibfield  {author} {\bibinfo {author} {\bibfnamefont {F.}~\bibnamefont {Family}}\ and\ \bibinfo {author} {\bibfnamefont {T.}~\bibnamefont {Vicsek}},\ }\bibfield  {title} {\bibinfo {title} {{Scaling of the active zone in the Eden process on percolation networks and the ballistic deposition model}},\ }\href {https://doi.org/10.1088/0305-4470/18/2/005} {\bibfield  {journal} {\bibinfo  {journal} {J. Phys. A: Math. Gen.}\ }\textbf {\bibinfo {volume} {18}},\ \bibinfo {pages} {L75} (\bibinfo {year} {1985})}\BibitemShut {NoStop}%
\bibitem [{\citenamefont {Hoshen}\ and\ \citenamefont {Kopelman}(1976)}]{Hoshen76}%
  \BibitemOpen
  \bibfield  {author} {\bibinfo {author} {\bibfnamefont {J.}~\bibnamefont {Hoshen}}\ and\ \bibinfo {author} {\bibfnamefont {R.}~\bibnamefont {Kopelman}},\ }\bibfield  {title} {\bibinfo {title} {{Percolation and cluster distribution. I. Cluster multiple labeling technique and critical concentration algorithm}},\ }\href {https://doi.org/10.1103/PhysRevB.14.3438} {\bibfield  {journal} {\bibinfo  {journal} {Phys. Rev. B}\ }\textbf {\bibinfo {volume} {14}},\ \bibinfo {pages} {3438} (\bibinfo {year} {1976})}\BibitemShut {NoStop}%
\bibitem [{\citenamefont {Corberi}\ \emph {et~al.}()\citenamefont {Corberi}, \citenamefont {Cugliandolo},\ and\ \citenamefont {Yoshino}}]{Corberi11}%
  \BibitemOpen
  \bibfield  {author} {\bibinfo {author} {\bibfnamefont {F.}~\bibnamefont {Corberi}}, \bibinfo {author} {\bibfnamefont {L.~F.}\ \bibnamefont {Cugliandolo}},\ and\ \bibinfo {author} {\bibfnamefont {H.}~\bibnamefont {Yoshino}},\ }\href@noop {} {\bibinfo {title} {Growing length scales in aging, in ``{D}ynamical heterogeneities in glasses, colloids, and granular media'', {L.} {B}erthier, {G}. {B}iroli, {J-P} {B}ouchaud, {L.} {C}ipelletti and {W.} van {S}aarloos eds. ({O}xford {U}niversity {P}ress, 2011)}}\BibitemShut {NoStop}%
\bibitem [{\citenamefont {Pagnani}\ and\ \citenamefont {Parisi}(2015)}]{Pagnani15}%
  \BibitemOpen
  \bibfield  {author} {\bibinfo {author} {\bibfnamefont {A.}~\bibnamefont {Pagnani}}\ and\ \bibinfo {author} {\bibfnamefont {G.}~\bibnamefont {Parisi}},\ }\bibfield  {title} {\bibinfo {title} {Numerical estimate of the {K}ardar-{P}arisi-{Z}hang universality class in (2+1) dimensions},\ }\href {https://doi.org/10.1103/PhysRevE.92.010101} {\bibfield  {journal} {\bibinfo  {journal} {Phys. Rev. E}\ }\textbf {\bibinfo {volume} {92}},\ \bibinfo {pages} {010101} (\bibinfo {year} {2015})}\BibitemShut {NoStop}%
\bibitem [{\citenamefont {Blanchard}\ \emph {et~al.}(2014)\citenamefont {Blanchard}, \citenamefont {Corberi}, \citenamefont {Cugliandolo},\ and\ \citenamefont {Picco}}]{Blanchard14}%
  \BibitemOpen
  \bibfield  {author} {\bibinfo {author} {\bibfnamefont {T.}~\bibnamefont {Blanchard}}, \bibinfo {author} {\bibfnamefont {F.}~\bibnamefont {Corberi}}, \bibinfo {author} {\bibfnamefont {L.~F.}\ \bibnamefont {Cugliandolo}},\ and\ \bibinfo {author} {\bibfnamefont {M.}~\bibnamefont {Picco}},\ }\bibfield  {title} {\bibinfo {title} {{How soon after a zero-temperature quench is the fate of the Ising model sealed?}},\ }\href {https://doi.org/10.1209/0295-5075/106/66001} {\bibfield  {journal} {\bibinfo  {journal} {Europhys. Lett.}\ }\textbf {\bibinfo {volume} {106}},\ \bibinfo {pages} {66001} (\bibinfo {year} {2014})}\BibitemShut {NoStop}%
\bibitem [{\citenamefont {Tartaglia}\ \emph {et~al.}(2015)\citenamefont {Tartaglia}, \citenamefont {Cugliandolo},\ and\ \citenamefont {Picco}}]{Tartaglia15}%
  \BibitemOpen
  \bibfield  {author} {\bibinfo {author} {\bibfnamefont {A.}~\bibnamefont {Tartaglia}}, \bibinfo {author} {\bibfnamefont {L.~F.}\ \bibnamefont {Cugliandolo}},\ and\ \bibinfo {author} {\bibfnamefont {M.}~\bibnamefont {Picco}},\ }\bibfield  {title} {\bibinfo {title} {Percolation and coarsening in the bidimensional voter model},\ }\href {https://doi.org/10.1103/PhysRevE.92.042109} {\bibfield  {journal} {\bibinfo  {journal} {Phys. Rev. E}\ }\textbf {\bibinfo {volume} {92}},\ \bibinfo {pages} {042109} (\bibinfo {year} {2015})}\BibitemShut {NoStop}%
\bibitem [{\citenamefont {Blanchard}\ \emph {et~al.}(2017)\citenamefont {Blanchard}, \citenamefont {Cugliandolo}, \citenamefont {Picco},\ and\ \citenamefont {Tartaglia}}]{Blanchard17}%
  \BibitemOpen
  \bibfield  {author} {\bibinfo {author} {\bibfnamefont {T.}~\bibnamefont {Blanchard}}, \bibinfo {author} {\bibfnamefont {L.~F.}\ \bibnamefont {Cugliandolo}}, \bibinfo {author} {\bibfnamefont {M.}~\bibnamefont {Picco}},\ and\ \bibinfo {author} {\bibfnamefont {A.}~\bibnamefont {Tartaglia}},\ }\bibfield  {title} {\bibinfo {title} {{Critical percolation in the dynamics of the 2D ferromagnetic Ising model}},\ }\href {https://doi.org/10.1088/1742-5468/aa9348} {\bibfield  {journal} {\bibinfo  {journal} {J. Stat. Mech.}\ }\textbf {\bibinfo {volume} {2017}},\ \bibinfo {pages} {113201} (\bibinfo {year} {2017})}\BibitemShut {NoStop}%
\bibitem [{\citenamefont {Tartaglia}\ \emph {et~al.}(2018)\citenamefont {Tartaglia}, \citenamefont {Cugliandolo},\ and\ \citenamefont {Picco}}]{Tartaglia18}%
  \BibitemOpen
  \bibfield  {author} {\bibinfo {author} {\bibfnamefont {A.}~\bibnamefont {Tartaglia}}, \bibinfo {author} {\bibfnamefont {L.~F.}\ \bibnamefont {Cugliandolo}},\ and\ \bibinfo {author} {\bibfnamefont {M.}~\bibnamefont {Picco}},\ }\bibfield  {title} {\bibinfo {title} {{Coarsening and percolation in the kinetic 2d Ising model with spin exchange updates and the voter model}},\ }\href {https://doi.org/10.1088/1742-5468/aad366} {\bibfield  {journal} {\bibinfo  {journal} {J. Stat. Mech.}\ }\textbf {\bibinfo {volume} {2018}},\ \bibinfo {pages} {083202} (\bibinfo {year} {2018})}\BibitemShut {NoStop}%
\bibitem [{\citenamefont {Blanchard}\ \emph {et~al.}(2012)\citenamefont {Blanchard}, \citenamefont {Cugliandolo},\ and\ \citenamefont {Picco}}]{Blanchard12}%
  \BibitemOpen
  \bibfield  {author} {\bibinfo {author} {\bibfnamefont {T.}~\bibnamefont {Blanchard}}, \bibinfo {author} {\bibfnamefont {L.~F.}\ \bibnamefont {Cugliandolo}},\ and\ \bibinfo {author} {\bibfnamefont {M.}~\bibnamefont {Picco}},\ }\bibfield  {title} {\bibinfo {title} {{A morphological study of cluster dynamics between critical points}},\ }\href {https://doi.org/10.1088/1742-5468/2012/05/P05026} {\bibfield  {journal} {\bibinfo  {journal} {J. Stat. Mech.}\ }\textbf {\bibinfo {volume} {2012}},\ \bibinfo {pages} {P05026} (\bibinfo {year} {2012})}\BibitemShut {NoStop}%
\bibitem [{Note2()}]{Note2}%
  \BibitemOpen
  \bibinfo {note} {Although the data are compatible with an algebraic regime ruled by $\tau _2$, it is not fully resolved yet whether, due to finite-size effects, this is an artifact of a regime preceding the exponential decay.}\BibitemShut {Stop}%
\bibitem [{Note3()}]{Note3}%
  \BibitemOpen
  \bibinfo {note} {Remarkably, this value is very close to the exponent of the distribution of holes within the critical percolation backbone, $\tau _b\simeq 1.82$~\cite {HuZiDe16}.}\BibitemShut {Stop}%
\bibitem [{\citenamefont {Duplantier}(1990)}]{Duplantier90}%
  \BibitemOpen
  \bibfield  {author} {\bibinfo {author} {\bibfnamefont {B.}~\bibnamefont {Duplantier}},\ }\bibfield  {title} {\bibinfo {title} {Exact fractal area of two-dimensional vesicles},\ }\href {https://doi.org/10.1103/PhysRevLett.64.493} {\bibfield  {journal} {\bibinfo  {journal} {Phys. Rev. Lett.}\ }\textbf {\bibinfo {volume} {64}},\ \bibinfo {pages} {493} (\bibinfo {year} {1990})}\BibitemShut {NoStop}%
\bibitem [{\citenamefont {Doherty}\ \emph {et~al.}(1994)\citenamefont {Doherty}, \citenamefont {Moore}, \citenamefont {Kim},\ and\ \citenamefont {Bray}}]{Doherty94}%
  \BibitemOpen
  \bibfield  {author} {\bibinfo {author} {\bibfnamefont {J.~P.}\ \bibnamefont {Doherty}}, \bibinfo {author} {\bibfnamefont {M.~A.}\ \bibnamefont {Moore}}, \bibinfo {author} {\bibfnamefont {J.~M.}\ \bibnamefont {Kim}},\ and\ \bibinfo {author} {\bibfnamefont {A.~J.}\ \bibnamefont {Bray}},\ }\bibfield  {title} {\bibinfo {title} {Generalizations of the {K}ardar-{P}arisi-{Z}hang equation},\ }\href {https://doi.org/10.1103/PhysRevLett.72.2041} {\bibfield  {journal} {\bibinfo  {journal} {Phys. Rev. Lett.}\ }\textbf {\bibinfo {volume} {72}},\ \bibinfo {pages} {2041} (\bibinfo {year} {1994})}\BibitemShut {NoStop}%
\bibitem [{\citenamefont {Tu}(1994)}]{Tu94}%
  \BibitemOpen
  \bibfield  {author} {\bibinfo {author} {\bibfnamefont {Y.}~\bibnamefont {Tu}},\ }\bibfield  {title} {\bibinfo {title} {Absence of finite upper critical dimension in the spherical {K}ardar-{P}arisi-{Z}hang model},\ }\href {https://doi.org/10.1103/PhysRevLett.73.3109} {\bibfield  {journal} {\bibinfo  {journal} {Phys. Rev. Lett.}\ }\textbf {\bibinfo {volume} {73}},\ \bibinfo {pages} {3109} (\bibinfo {year} {1994})}\BibitemShut {NoStop}%
\bibitem [{\citenamefont {Colaiori}\ and\ \citenamefont {Moore}(2001)}]{Colaiori01}%
  \BibitemOpen
  \bibfield  {author} {\bibinfo {author} {\bibfnamefont {F.}~\bibnamefont {Colaiori}}\ and\ \bibinfo {author} {\bibfnamefont {M.~A.}\ \bibnamefont {Moore}},\ }\bibfield  {title} {\bibinfo {title} {Upper critical dimension, dynamic exponent, and scaling functions in the mode-coupling theory for the {K}ardar-{P}arisi-{Z}hang equation},\ }\href {https://doi.org/10.1103/PhysRevLett.86.3946} {\bibfield  {journal} {\bibinfo  {journal} {Phys. Rev. Lett.}\ }\textbf {\bibinfo {volume} {86}},\ \bibinfo {pages} {3946} (\bibinfo {year} {2001})}\BibitemShut {NoStop}%
\bibitem [{\citenamefont {L\"assig}(1998)}]{Lassig98}%
  \BibitemOpen
  \bibfield  {author} {\bibinfo {author} {\bibfnamefont {M.}~\bibnamefont {L\"assig}},\ }\bibfield  {title} {\bibinfo {title} {Quantized scaling of growing surfaces},\ }\href {https://doi.org/10.1103/PhysRevLett.80.2366} {\bibfield  {journal} {\bibinfo  {journal} {Phys. Rev. Lett.}\ }\textbf {\bibinfo {volume} {80}},\ \bibinfo {pages} {2366} (\bibinfo {year} {1998})}\BibitemShut {NoStop}%
\bibitem [{\citenamefont {Canet}\ \emph {et~al.}(2010)\citenamefont {Canet}, \citenamefont {Chat\'e}, \citenamefont {Delamotte},\ and\ \citenamefont {Wschebor}}]{Canet10}%
  \BibitemOpen
  \bibfield  {author} {\bibinfo {author} {\bibfnamefont {L.}~\bibnamefont {Canet}}, \bibinfo {author} {\bibfnamefont {H.}~\bibnamefont {Chat\'e}}, \bibinfo {author} {\bibfnamefont {B.}~\bibnamefont {Delamotte}},\ and\ \bibinfo {author} {\bibfnamefont {N.}~\bibnamefont {Wschebor}},\ }\bibfield  {title} {\bibinfo {title} {Nonperturbative renormalization group for the {K}ardar-{P}arisi-{Z}hang equation},\ }\href {https://doi.org/10.1103/PhysRevLett.104.150601} {\bibfield  {journal} {\bibinfo  {journal} {Phys. Rev. Lett.}\ }\textbf {\bibinfo {volume} {104}},\ \bibinfo {pages} {150601} (\bibinfo {year} {2010})}\BibitemShut {NoStop}%
\bibitem [{\citenamefont {Kloss}\ \emph {et~al.}(2012)\citenamefont {Kloss}, \citenamefont {Canet},\ and\ \citenamefont {Wschebor}}]{Kloss12}%
  \BibitemOpen
  \bibfield  {author} {\bibinfo {author} {\bibfnamefont {T.}~\bibnamefont {Kloss}}, \bibinfo {author} {\bibfnamefont {L.}~\bibnamefont {Canet}},\ and\ \bibinfo {author} {\bibfnamefont {N.}~\bibnamefont {Wschebor}},\ }\bibfield  {title} {\bibinfo {title} {Nonperturbative renormalization group for the stationary {K}ardar-{P}arisi-{Z}hang equation: Scaling functions and amplitude ratios in 1+1, 2+1, and 3+1 dimensions},\ }\href {https://doi.org/10.1103/PhysRevE.86.051124} {\bibfield  {journal} {\bibinfo  {journal} {Phys. Rev. E}\ }\textbf {\bibinfo {volume} {86}},\ \bibinfo {pages} {051124} (\bibinfo {year} {2012})}\BibitemShut {NoStop}%
\bibitem [{\citenamefont {Wolf}\ and\ \citenamefont {Kertész}(1987)}]{Wolf87}%
  \BibitemOpen
  \bibfield  {author} {\bibinfo {author} {\bibfnamefont {D.~E.}\ \bibnamefont {Wolf}}\ and\ \bibinfo {author} {\bibfnamefont {J.}~\bibnamefont {Kertész}},\ }\bibfield  {title} {\bibinfo {title} {Surface width exponents for three- and four-dimensional {E}den growth},\ }\href {https://doi.org/10.1209/0295-5075/4/6/003} {\bibfield  {journal} {\bibinfo  {journal} {Europhys. Lett.}\ }\textbf {\bibinfo {volume} {4}},\ \bibinfo {pages} {651} (\bibinfo {year} {1987})}\BibitemShut {NoStop}%
\bibitem [{\citenamefont {Oliveira}(2022)}]{Oliveira22}%
  \BibitemOpen
  \bibfield  {author} {\bibinfo {author} {\bibfnamefont {T.~J.}\ \bibnamefont {Oliveira}},\ }\bibfield  {title} {\bibinfo {title} {{K}ardar-{P}arisi-{Z}hang universality class in $(d+1)$-dimensions},\ }\href {https://doi.org/10.1103/PhysRevE.106.L062103} {\bibfield  {journal} {\bibinfo  {journal} {Phys. Rev. E}\ }\textbf {\bibinfo {volume} {106}},\ \bibinfo {pages} {L062103} (\bibinfo {year} {2022})}\BibitemShut {NoStop}%
\bibitem [{\citenamefont {Kelling}\ and\ \citenamefont {\'Odor}(2011)}]{Kelling11}%
  \BibitemOpen
  \bibfield  {author} {\bibinfo {author} {\bibfnamefont {J.}~\bibnamefont {Kelling}}\ and\ \bibinfo {author} {\bibfnamefont {G.}~\bibnamefont {\'Odor}},\ }\bibfield  {title} {\bibinfo {title} {Extremely large-scale simulation of a {K}ardar-{P}arisi-{Z}hang model using graphics cards},\ }\href {https://doi.org/10.1103/PhysRevE.84.061150} {\bibfield  {journal} {\bibinfo  {journal} {Phys. Rev. E}\ }\textbf {\bibinfo {volume} {84}},\ \bibinfo {pages} {061150} (\bibinfo {year} {2011})}\BibitemShut {NoStop}%
\bibitem [{\citenamefont {Kelling}\ \emph {et~al.}(2017)\citenamefont {Kelling}, \citenamefont {Ódor},\ and\ \citenamefont {Gemming}}]{Kelling18}%
  \BibitemOpen
  \bibfield  {author} {\bibinfo {author} {\bibfnamefont {J.}~\bibnamefont {Kelling}}, \bibinfo {author} {\bibfnamefont {G.}~\bibnamefont {Ódor}},\ and\ \bibinfo {author} {\bibfnamefont {S.}~\bibnamefont {Gemming}},\ }\bibfield  {title} {\bibinfo {title} {Dynamical universality classes of simple growth and lattice gas models},\ }\href {https://doi.org/10.1088/1751-8121/aa97f3} {\bibfield  {journal} {\bibinfo  {journal} {J. Phys. A: Math. Theor.}\ }\textbf {\bibinfo {volume} {51}},\ \bibinfo {pages} {035003} (\bibinfo {year} {2017})}\BibitemShut {NoStop}%
\bibitem [{\citenamefont {Hu}\ \emph {et~al.}(2016)\citenamefont {Hu}, \citenamefont {Ziff},\ and\ \citenamefont {Deng}}]{HuZiDe16}%
  \BibitemOpen
  \bibfield  {author} {\bibinfo {author} {\bibfnamefont {H.}~\bibnamefont {Hu}}, \bibinfo {author} {\bibfnamefont {R.~M.}\ \bibnamefont {Ziff}},\ and\ \bibinfo {author} {\bibfnamefont {Y.}~\bibnamefont {Deng}},\ }\bibfield  {title} {\bibinfo {title} {{No-Enclave Percolation Corresponds to Holes in the Cluster Backbone}},\ }\href {https://doi.org/10.1103/PhysRevLett.117.185701} {\bibfield  {journal} {\bibinfo  {journal} {Phys. Rev. Lett.}\ }\textbf {\bibinfo {volume} {117}},\ \bibinfo {pages} {185701} (\bibinfo {year} {2016})}\BibitemShut {NoStop}%
\end{thebibliography}%

\end{document}